\documentclass{aa}
\usepackage[varg]{txfonts}

\usepackage{graphicx}
\usepackage{tikz}
\usepackage[breaklinks=true]{hyperref}
\bibpunct{(}{)}{;}{a}{}{,} % to follow the A&A style

\renewcommand{\a}{\ensuremath{\mathbf{a}}}
\newcommand{\rhat}{\ensuremath{\hat{\mathbf{r}}}}
\newcommand{\rh}{\ensuremath{r_\mathrm{h}}}
\newcommand{\au}{\ensuremath{\,\mathrm{au}}}
\newcommand{\um}{\ensuremath{\,\mathrm{\mu m}}}
\newcommand{\cm}{\ensuremath{\,\mathrm{cm}}}
\newcommand{\km}{\ensuremath{\,\mathrm{km}}}
\newcommand{\h}{\ensuremath{\,\mathrm{h}}}
\newcommand{\den}{\ensuremath{\,\mathrm{kg/m^3}}}
\newcommand{\tiu}{\ensuremath{\,\mathrm{tiu}}}

\begin{document}

\title{Thermal radiation pressure as a possible mechanism for losing small particles on asteroids}

\author{Yoonsoo P. Bach\inst{1, 2}
  \and Masateru Ishiguro\inst{1, 2}}

\institute{
  Department of Physics and Astronomy, Seoul National University, Gwanak-ro 1, Gwanak-gu, Seoul 08826, Republic of Korea\and
  SNU Astronomy Research Center, Department of Physics and Astronomy, Seoul National University, Gwanak-ro 1, Gwanak-gu, Seoul 08826, Republic of Korea\\
  \email{ishiguro@astro.snu.ac.kr, ysbach93@gmail.com}}

   \date{Received ; accepted }

\abstract 
{Recent observations of dust ejections from active asteroids, including (3200) Phaethon, have drawn considerable interest from planetary astronomers studying the generation and removal of small dust particles on asteroids.} 
{In this work, we aim to investigate the importance of thermal radiation pressure from asteroid regolith (AR) acting on small dust particles over the surface of the AR. In particular, we aim to understand the role of thermal radiation in the near-Sun environment.} 
{We describe the acceleration of particles over the AR within the radiation fields (direct solar, reflected (scattered) solar, and thermal radiation) in addition to the asteroid's rotation and gravitational field. Mie theory is used because the particles of interest have sizes comparable to thermal wavelengths ($ \sim 1\mathrm{-}100 \um $), and thus the geometric approximation is not applicable. A new set of formalisms is developed for the purpose.} 
{We find that the acceleration of particles with spherical radius $ \lesssim 1\um $ to $ \sim 10 \um $ is dominated by the thermal radiation from the AR when the asteroid is in the near-Sun environment (heliocentric distance $ \rh \lesssim 0.8 \au $). Under thermal radiation dominance, the net acceleration is towards space, that is, outwards from the AR. This outward acceleration is the strongest for particles of $ \sim 1 \um $ in radius, regardless of other parameters. A preliminary trajectory integration using the Phaethon-like model shows that such particles escape from the gravitational field within about 10 minutes. Our results are consistent with the previous observational studies on Phaethon in that the ejected dust particles have a spherical radius of $ \sim 1\um $.} 
{}


\titlerunning{Thermal radiation pressure on asteroids}
\maketitle

%%%%%%%%%%%%%%%%%%%%%%%%%%%%%%%%%%%%%%%%%%%%%%%%%%%%%%%%%%%%%%%%%%%%%%%%%%%%%%%%
\section{Introduction}
%%%%%%%%%%%%%%%%%%%%%%%%%%%%%%%%%%%%%%%%%%%%%%%%%%%%%%%%%%%%%%%%%%%%%%%%%%%%%%%%

Recent discoveries of active asteroids \citep[reviewed in, e.g.,][]{2012AJ....143...66J} and the paucity of small dust particles on the Hayabusa2 target asteroid (162173) Ryugu \citep{2019NatAs...3..971G} have drawn considerable interest from planetary astronomers studying the generation and removal of small (submillimeter; subm(mm) dust particles.  Centimeter(cm)-sized particles were observed being ejected from the asteroid (101955) Bennu   \citep{2019Sci...366.3544L}, and later their orbits were physically modeled \citep{McMahon2020}. Also, the near-Sun asteroid (3200) Phaethon, the target body of the JAXA DESTINY$ ^+ $ mission \citep{2018LPI....49.2570A}, is a symbolic asteroid, as recurrent dust ejections have been witnessed through telescopic observations around its perihelion over the past decade \citep{2010AJ....140.1519J, 2013ApJ...771L..36J, 2017AJ....153...23H}. 

%$ \rh \sim 1 \au $ \citep{1985MNRAS.214P..29G, 1992Icar...97..276L, 1996Icar..119..173C, 2005ApJ...624.1093H, 2008Icar..194..843W, 2018AJ....156..238J, 2019AJ....157..193J}

From  observations, the effective radius of the  particles ejected from Phaethon is estimated to be as small as $ \sim 1 \um $ (\citealt{2013ApJ...771L..36J, 2017AJ....153...23H}). Although the generation mechanism of such small particles has been thoroughly studied in terms of thermal fracturing in an astronomical context \citep{2010AJ....140.1519J, 2014Natur.508..233D, 2017Icar..294..247M}, the mechanism responsible for accelerating and ejecting these dust particles around the perihelion remains unexplored.

% Will it necessary to discuss about the linkage to Geminids?
%Phaethon has long been suspected as the parent body of Geminids meteor stream \citep{1989A&A...225..533G, 1993MNRAS.262..231W}, but the estimated (episodic) mass production rate from these direct Phaethon observations is orders of magnitude smaller than the required ones from meteor showers \citep{2017P&SS..143...83B, 2017P&SS..143..125R} as discussed in \cite{2019AJ....157..193J}. Notwithstanding, it will be important to understand how the particles are ejected near the perihelion.

In this context, the first question to tackle is that regarding the unknown mechanism responsible for ejecting such particles from asteroids. The mechanisms on Bennu and Phaethon may likely differ because of the large size difference of the particles (cm and $ \mathrm{\mu m} $) and the fact that Phaethon showed recurrent activity near its perihelion. Phaethon is an important object because it is the only object with detectable recurrent dust ejection. There have been questions such as whether Phaethon's dust ejections near perihelion can explain the total mass of the Geminid meteor stream \citep{2017P&SS..143...83B, 2017P&SS..143..125R, 2019AJ....157..193J}. Such questions can be discussed more systematically once the dust ejection mechanisms are known. 
%Further discussions to find a way to explain such mass production inconsistency can be fruitful after understanding the physics of this activity. 
The aim of this work is to find a mechanism able to explain many of the observations of Phaethon's activity, at least to some degree.

When an asteroid comes close to the Sun, the surface is heated by solar radiation. At a heliocentric distance of $ 0.2 \au $, the maximum surface temperature reaches $ \sim 900 \,\mathrm{K} $. Along the terminator on a near-Sun asteroid, the rapid heating or cooling may induce severe stress on the surface material \citep{2017Icar..294..247M}, which will lead to thermal fracturing, as demonstrated in laboratory experiments \citep{2014Natur.508..233D}. Accordingly, {small} dust particles may be generated on the asteroid surface \citep{2010AJ....140.1519J}. 
% Some description about terrestrial environment (removed)
%The stress induced by thermal expansion is sufficient to produce large-scale fractures or exfoliation fractures on terrestrial rocks with subcritical growth, even without precipitation, freezing, or seismic activity \citep{2016NatGe...9..395C}. Explosive exfoliation is sometimes observed \citep{2018NatCo...9..762C}. However, such explosive exfoliation is possible only if tectonic stress is added to the thermal stress, which is not possible on asteroids. Stress may also be induced by mineral desiccation at high surface temperatures \cite{2012AJ....143...66J}.
%Recent thermal modeling on Phaethon also confirmed that its surface temperature variation is likely to exert enough thermal stress to cause thermal fatigue (\cite{2019MNRAS.482.4243Y}), and therefore may supply small debris on the surface.
Meteorites are the most direct sources with which  to infer the size of such small particles. Studies have found meteoritic matrix particles with sizes down to $ 0.01 \mathrm{-} 5 \um $ \citep{2005ApJ...623..571S, 2006mess.book..803R}, which suggests that dust particles of $ \sim 1 \um $ or smaller could be present on asteroids.

% Here, we propose that the acceleration of the thus generated dust particles is affected by thermal radiation from the scorching asteroid regolith (AR) surface in the near-Sun environment. Although the thermal radiation of the Sun is more significant than that of a small patch on the AR surface, the latter must be integrated over the AR surface that contributes the radiation to the particle. It is not trivial, therefore, to state the latter is always negligible. In this study, we confirm the radiative acceleration from the asteroid surface can be the dominant acceleration term, and the thermal radiation can make dust particles overcome the solar radiation and asteroid's gravity under certain circumstances. It is worth noting that our research successfully explains the observed dust ejection phenomenon from Phaethon for the first time. Besides, we provide a \textit{testable prediction} about the ejecta size range once thermal fracture happens on general asteroids.

Applications of thermal radiation pressure in similar contexts are presented in \cite{McMahon2020} (see \citealt{2019Sci...366.3544L} for the activity of   Bennu) and the spacecraft control \citep{2017JGCD...40.2432H}. However, the formalisms in those works are applicable only to large particles (geometric regime) at large initial height (typically around $ 10 \,\mathrm{m} $). Using the formalism, \cite{McMahon2020} found that particles with initial speed $ < 10 \,\mathrm{cm/s} $ cannot escape from Bennu. 

% We insist that these models are not satisfactory for explaining the dust ejection from near-Sun asteroids sufficiently because of the reason below. First, these works do not take the scattering theory of small particles into account but used a geometric approximation. For particles much smaller than mm scale are not manageable by the approximated model, \cite{McMahon2020} investigated large particles as observed \citep{2019Sci...366.3544L} with sizes down to $ 1 \,\mathrm{mm} $. Furthermore, the formalism is applicable only if particles are at a height greater than about $ 10 \,\mathrm{m} $ from the AR (greater than the length scale of the shape model). In addition, it is mentioned that no particle with initial speed below $ 10 \,\mathrm{cm/s} $ could escape from Bennu \citep{McMahon2020}, although it is unclear what induces this initial thrust. Lastly, there is no further discussion about any other environmental factors, such as heliocentric distance or asteroid's physical properties.

In this work, we consider small stationary particles (radius $ 0.5 $ to tens of microns) at a small height of the order of $ 1 \cm$ above the asteroid regolith (AR) without any initial speed. Our results demonstrate that neither initial thrust nor large initial height are essential for such small particles to be accelerated and escape into space in the near-Sun environment. 
% Removed in review #1
%Therefore, from the motivation to formalism, previous works such as \cite{McMahon2020} are not on the same line with this work. We contrived this idea and developed the formalism provided here well before the publication of \cite{McMahon2020}, without knowing any ongoing work treating thermal radiation pressure acting on particles in astronomical contexts, devising a set of formalism largely different from it \citep{Bach+Ishiguro_IDP2019}.

%%%%%%%%%%%%%%%%%%%%%%%%%%%%%%%%%%%%%%%%%%%%%%%%%%%%%%%%%%%%%%%%%%%%%%%%%%%%%%%%
\section{Methods} \label{meth}
%%%%%%%%%%%%%%%%%%%%%%%%%%%%%%%%%%%%%%%%%%%%%%%%%%%%%%%%%%%%%%%%%%%%%%%%%%%%%%%%

A particle floating above the AR experiences the following forces: direct solar radiation, indirect solar radiation reflected (scattered) on AR, thermal radiation from AR, and the gravity of the asteroid. The acceleration vectors associated to these are denoted $ \a_\mathrm{\odot} $, $ \a_\mathrm{ref} $, $ \a_\mathrm{ther} $, and $ \a_\mathrm{grav} $, respectively. In addition to gravity, the centrifugal term ($ \a_\mathrm{cen} $) is included if we choose a non-inertial rotating frame on the asteroid. The net \textit{radiative} acceleration is defined such that $ \a_\mathrm{rad} = \a_\odot + \a_\mathrm{ther} + \a_\mathrm{ref} $. The net acceleration is the sum of all accelerations, $ \a_\mathrm{total} = \a_\mathrm{rad} + (\a_\mathrm{grav} + \a_\mathrm{cen}) $. Here, $ \a_\mathrm{ther} $ and $ \a_\mathrm{ref} $ are directed outward from the AR, while $ \a_\odot $ is usually directed inward. 
% a_cen included = non-inertial rotating frame // a_cen not included = inertial frame.
The timescale of interest, for example the time required for a particle to leave the asteroid's gravitational field, is assumed to be short compared to the asteroid's orbital timescale, and hence the orbital motion of the  asteroid is not considered. 
% Removed during the 1st reviewing process (2021-02-22 16:18:56 (KST: GMT+09:00) YPB)
%The acceleration by the solar tidal force is not considered since the asteroid size divided by the heliocentric distance is too small ($ D/\rh \ll 1 $).  

\subsection{Problem statement}\label{meth-problem statement}
%%%%%%%%%%%%%%%%%%%%%%%%%%%%%%%%%%%%%%%%%%%%%%%%%%%%%%%%%%%%%%%%%%%%%%%%%%%%%%%%
In this work, we aim to calculate the net acceleration acting on a particle that is placed above the AR at an initial height of $ H $. The way in which we achieve this initial condition is discussed in Sect. \ref{initial condition}; it is taken as granted for the modeling in this work. To physically handle this problem with manageable complexity, the following assumptions are used.

The particle is a homogeneous sphere of radius $ r_a $ with a known refractive index. The condition of height, $ H \gg r_a $, should be satisfied all the time. Then any radiations from AR can be approximated as a plane wave (similar to solar irradiation), ignoring the curvature of the particle. Therefore, the Mie theory (Lorenz--Mie theory) is applicable.

Under the Mie scattering theory \citep[see][chapter 4]{1998asls.book.....B}, the effective cross section for radiative pressure is calculated for a given particle size, refractive index, and wavelength (Sects. \ref{meth-mie}). If the spectrum of the incident radiation is known, the radiation pressure and acceleration are calculated by integrating the contribution from all wavelengths.

Throughout this work, AR is assumed as a Lambertian scatterer. Although more comprehensive models are available \citep[e.g.,][]{hapke2012book}, they would introduce multiple new free parameters, while the Lambertian scattering is a good approximation to the complicated scattering process. 
% erased - too detailed
% Moreover, for typical Bond albedos studied in this work ($ A_B \le 0.05 $; Sect. \ref{meth-model param selection} and Table \ref{tab: phys_phae}), the acceleration by scattered light will be negligible to that by the solar irradiation: $ \| \mathbf{a}_\mathrm{ref} \| \sim A_B \| \mathbf{a}_\odot \| $ (Eq. \ref{eq: acc_ref}) . 

The calculation of thermal radiation from AR requires the temperature distribution on the AR to be known a priori. For this, a smooth spherical asteroid with a 1D thermal conduction model is adopted to calculate the temperature of the  AR (see Sect. \ref{meth-tpm}). 

% Then the AR is assumed planar and isothermal in the particle's field of view. This assumption should be used with care when the height $ H $ gets comparable or larger than the ``isothermal length scale'' on the AR. For simplicity, the isothermal patch on the AR is assumed to be circular with radius $ r_0 $, $ r_a \ll H \ll r_0 $, while $ r_0 \ll D $, where $ D $ is the diameter of the asteroid. $ r_0 $ is fixed to $ 0.005 D $ (see Appendix \ref{app-r_0}).

By combining the assumptions mentioned above, a simple set of descriptions of forces is established. In the following sections, we describe how the AR surface temperature $ T_S $ is obtained using thermal modeling, how the Mie theory is applied to calculate the radiative accelerations, and a detailed set of derivations of the formalisms used throughout this work. Justifications of selected parameter values used in this work are then provided. Finally, we outline a simple way to compare the magnitudes of solar and thermal radiations based on the developed formalism.

\subsection{Thermal modeling} \label{meth-tpm}
%%%%%%%%%%%%%%%%%%%%%%%%%%%%%%%%%%%%%%%%%%%%%%%%%%%%%%%%%%%%%%%%%%%%%%%%%%%%%%%%
The surface temperature of an asteroid is necessary for estimating $ \a_\mathrm{ther} $. The surface temperature is calculated by the so-called thermo-physical model (TPM). In this section, we describe the outline of the widely used scheme \citep{1989Icar...78..337S, DelboM2004PhDT, MuellerM2007PhDT} adopted in this work. A reparameterization of the problem is then summarized; this is used throughout the present work.

In the scheme, the following classic 1D thermal conduction equation (Fourier's equation) on the surface is solved under proper boundary conditions:
\begin{equation}\label{eq:tpm}
  \rho c_s \frac{\partial T}{\partial t} = \kappa \frac{\partial^2 T}{\partial z^2} ~.
\end{equation}
Here, $ \rho $, $ c_s $, and $ \kappa $ denote the material mass density, the specific heat at constant pressure, and the thermal conductivity, respectively. $ T $ is the temperature of the  topmost cell in the simulation (we refer to this as the AR \textit{surface} temperature), $ t $ is the time, and $ z $ is the depth \citep{1989Icar...78..337S, MuellerM2007PhDT}. This conduction equation is solved by two boundary conditions: (i) the energy at the AR surface is balanced between the incident solar radiation energy, the subsurface conduction energy, and the radiative emission energy; and (ii) the temperature gradient vanishes at infinite depth ($ \lim_{z \rightarrow \infty} \partial T(z, t)/\partial z = 0 $). The first boundary condition (i) is solved via the forward time-centered space scheme following the above-mentioned references. The emissivity used in (i) must match with the emissivity $ \epsilon_S(\lambda) $ to be used in the $ \a_\mathrm{ther} $ calculation below. The ``infinite depth'' in the second condition (ii) is set to ten times the thermal skin depth $ l_S = \sqrt{ \kappa P_\mathrm{rot} / (2\pi \rho c_s)} $, where $ P_\mathrm{rot} $ is the rotational period of the asteroid \citep[following, e.g.,][]{1989Icar...78..337S,MuellerM2007PhDT}. The Sun is assumed as a point source, although it can be as large as $ \sim 2.5^\circ $ at $ 0.2 \au $.

The modeled asteroid is assumed to be a smooth sphere; that is, any roughness and large-scale geological structures are ignored. There are a few advantages of ignoring roughness. First, the model is mostly described by the thermal parameter (Eq. (\ref{eq: theta-def})) except for the heliocentric distance and pole orientation. Thus, the discussions become analytically simple. An example is the effect of the rotational period and thermal inertia, as described below Eq. (\ref{eq: theta-def}). Second, a direct comparison with other theoretical studies such as \cite{2019JGRE..124.3304R} is possible as they also used smooth spherical asteroids (see Sect. \ref{initial condition}). Consideration of roughness (self-heating, shadowing, and/or 3D conduction) is computationally demanding and increases the dimensionality of the parameter space enormously.

The rotational period $ P_\mathrm{rot} $ is assumed to be significantly shorter than the orbital one. Thermal equilibrium is therefore reached at each fixed heliocentric distance, and the seasonal effect is neglected. To calculate the AR surface temperature, an initial condition of $ T(z) = T_\mathrm{eqm} e^{-z/l_S} $ is set for depth $ z $ at the local noon for a given latitude, where $ T_\mathrm{eqm} $ is defined below in Eq (\ref{eq: T_eqm}). This array is then evolved for $ N_\mathrm{iter}^\mathrm{min} = 50 $ rotations, and the rotation is further repeated until either the specified temperature stability of $ \sim 10^{-5} \,\mathrm{K} $ is reached at the surface or the iteration reaches 5000. It is confirmed that $ N_\mathrm{iter}^\mathrm{min} $ changes the surface temperature by $ \ll 0.1 \,\mathrm{K} $, given $ N_\mathrm{iter}^\mathrm{min} > 10 $. The resulting temperature along the longitude and depth during the last rotation becomes the temperature of the AR. If the Sun does not rise at a given latitude, the surface temperature is fixed to 0 K for all longitudes. The temperature is calculated for patches with a resolution of $ 1^\circ $ in both latitude and longitude. From these gridded values of the surface temperature, the temperature at an arbitrary position on the asteroid is calculated by linear interpolation (\texttt{RectBivariateSpline} of \texttt{scipy} without smoothing factor).

Equation (\ref{eq:tpm}) implies that there exists a set of effective $ \rho $, $ c_s$, and $\kappa $ ---and sometimes including the porosity factor--- that describes the realistic regolith. It is further assumed that $ \rho $, $ c_s $, and $ \kappa $, and thus thermal inertia ($ \Gamma := \sqrt{\kappa \rho c_s} $), are independent of temperature (see Sect. \ref{disc-effect of TI} for further discussion). The thermal inertia has the unit of $ \mathrm{tiu} \equiv \mathrm{J\, m^{-2}\, K^{-1}\, s^{-1/2}} $. The notation ``tiu'' stands for ``thermal inertia unit'', first coined by \cite{2006PhDT........15P}. 

All equations used in TPM are then finally solved by specifying the spin vector and two parameters. The first parameter is $ T_\mathrm{eqm} $, the equilibrium temperature at the subsolar point of a non-rotating asteroid or a plate: 
\begin{equation}\label{eq: T_eqm}
\begin{aligned}
  T_\mathrm{eqm} &= T_\mathrm{eqm} (A_B, \bar{\epsilon}_S, \rh) \\
    &= \left (
      \frac{1 - A_B}{\bar{\epsilon}_S \sigma_\mathrm{SB}} \frac{L_\odot}{4\pi \rh^2}
    \right )^{1/4}
    = T_1 c_T
      \left ( \frac{\rh}{1 \au} \right )^{-1/2}
    ~.
\end{aligned}
\end{equation}
The variables used in this equation are explained as follows. First,
\begin{equation}\label{eq: A_B}
  A_B \approx A_V \approx (0.286 + 0.656 G)p_\mathrm{V}
\end{equation}
is the Bond albedo of an asteroid. $ G $ is the slope parameter under the IAU $ H $, $ G $ magnitude system and $ p_\mathrm{V} $ is the geometric albedo in V-band (first definition by \citealt{1989aste.conf..524B}, corrected by \citealt{2016PASP..128d5004M}). Also,
\begin{equation}\label{eq: c_T-def}
   c_T := \left ( \frac{1-A_B} {\bar{\epsilon}_S } \right )^{1/4}
\end{equation}
is a constant for an AR model, and $ \bar{\epsilon}_S $ is the spectrum-averaged hemispherical emissivity under a blackbody spectrum of temperature $ T_S $, defined in Eq. (\ref{eq: epsilon-bar}). Conventionally, $ \bar{\epsilon}_S $ is fixed as a constant. It is fixed to 0.90 in this work. As $ A_B \ll 1 $ and $ \bar{\epsilon}_S \sim 1 $, the constant $ c_T \approx 1 $. The temperature
\begin{equation}\label{eq: T1-def}
  T_1 
    := \left ( \frac{L_\odot}{4\pi (1 \au)^2 \sigma_\mathrm{SB}} \right )^{1/4} 
    = 394.0 \,\mathrm{K} ~
\end{equation}
is the $ T_\mathrm{eqm} $ of a blackbody ($ \bar{\epsilon}_S = 0 $ and $ A_B = 0 $ from Kirchhoff's law) at $ 1 \au $.
The second parameter describing the TPM result is the thermal parameter \citep{1989Icar...78..337S,MuellerM2007PhDT}:
\begin{equation}\label{eq: theta-def}
\begin{aligned}
  \Theta 
    &= \Theta(\Gamma, P_\mathrm{rot}, A_B, \rh, \epsilon_S) \\
    &= \frac{\Gamma}{T_\mathrm{eqm}^3 \bar{\epsilon}_S \sigma_\mathrm{SB}}
      \sqrt{\frac{2\pi}{P_\mathrm{rot}}} \\
    &= \frac{1.2046}{\bar{\epsilon}_S c_T^3}
      \left ( \frac{\Gamma}{100 \tiu} \right )
      \left ( \frac{\rh}{1 \au} \right )^{3/2}
      \left ( \frac{P_\mathrm{rot}}{1 \h} \right )^{-1/2}
       ~.
\end{aligned}
\end{equation}

The TPM calculates the profile of a dimensionless quantity, $ T/T_\mathrm{eqm} $, on the AR over time (rotation). This profile is uniquely determined once the spin vector and $ \Theta $ are specified for the smooth model; converting it to the absolute temperature then requires $ T_\mathrm{eqm} $. An advantage of assuming a smooth, spherical asteroid is that $ \Theta $ is proportional to $ \Gamma / \sqrt{P_\mathrm{rot}} $. Therefore, changing $ P_\mathrm{rot} $ by a factor of $ 0.5 $, for example, is identical to changing $ \Gamma $ to $ \sqrt{2} \Gamma $ because both $ \Theta $ and $ T_\mathrm{eqm} $ are not changed by the modification (this fact is used in Sect. \ref{res-size-size}). Only the centrifugal acceleration must be updated according to $ P_\mathrm{rot} $.

%To wrap up, the surface temperature at given longitude and latitude is $ T_S = T_S(\mathrm{lon}, \mathrm{lat} \,|\, \Theta, \mathbf{S}) $ for spin direction $ \mathbf{S} $ .
%A further discussion on the suitability of some of our assumptions is given in the discussion section. 

\subsection{Mie calculation} \label{meth-mie}
%%%%%%%%%%%%%%%%%%%%%%%%%%%%%%%%%%%%%%%%%%%%%%%%%%%%%%%%%%%%%%%%%%%%%%%%%%%%%%%%
 Mie theory requires (A) a homogeneous spherical particle and (B) plane parallel incident electromagnetic waves \citep[see, e.g.,][]{1983asls.book.....B}. It then enables us to calculate the radiation pressure coefficient $ Q_\mathrm{pr} = Q_\mathrm{pr} (\lambda, r_a | \mathcal{I}_a ) $ once the refractive index is known. Here, $ \mathcal{I}_a $ stands for the information on the particle, the refractive index. The radiation pressure experienced by the particle is proportional to $ Q_\mathrm{pr}\sigma $, as well as the incident irradiance. If the extinction and scattering coefficients ($ Q_\mathrm{ext} $ and $ Q_\mathrm{sca} $, respectively) are given, then $ Q_\mathrm{pr} $ is calculated as 
\begin{equation}\label{eq: Q_pr def}
 Q_\mathrm{pr} = Q_\mathrm{ext} - Q_\mathrm{sca} \langle \cos \vartheta \rangle ~,
\end{equation}
where $ \vartheta $ is the scattering angle. The second term ($ Q_\mathrm{sca} $ multiplied by the so-called asymmetry factor, $ \langle \cos \vartheta \rangle $) takes only the momentum transferred towards the propagation direction (see e.g., Sect 2.3 of \citealt{1981lssp.book.....V} and Sect. 4.5 of \citealt{1983asls.book.....B}). Each term in the equation can be calculated once the particle radius $ r_a $ and refractive index are found using Mie theory. For large particles ($ r_a \gg \lambda $), a geometric limit is expected: $ Q_\mathrm{pr} \rightarrow 1 $.

For the optical indices, two well-studied materials are considered, namely olivine and magnetite. The former is transparent and the latter is opaque material in optical wavelengths. The optical indices are directly adopted from previous experimental works: low-Fe olivine ($\mathrm{Mg_{1.9} Fe_{0.1} Si O_4}$; \citealt{2001A&A...378..228F}\footnote{\label{foot: olivine}\url{https://www.astro.uni-jena.de/Laboratory/OCDB/newsilicates.html\#F}}) and magnetite ($ \mathrm{Fe_3 O_4} $; Triaud, A. H. M. J.\footnote{\label{foot: magnetite}\url{https://www.astro.uni-jena.de/Laboratory/OCDB/mgfeoxides.html}}; Mutschke, H. (2021), private comm.). Although they are not representative materials in the real AR compared to mixtures like meteoritic samples, our model can easily be extended to any real material with a known refractive index. These materials are chosen because the refractive indices are accurately known. The calculated $ Q_\mathrm{pr}(\lambda, r_a) $ values for these two materials are given in Appendix \ref{app: qpr}.

For the requirement (B) to hold, an assumption $ H \gg r_a $ is needed such that the solid angle of the particle viewed from an infinitesimal AR patch ($ \sim \sigma / H^2 = \pi r_a^2 / H^2 $) is much smaller than unity. As shown in Sect. 3, the $ r_a $ of interest does not exceed a few tens of microns, and hence an arbitrary choice of $ H = 1 \cm $ satisfies the requirement.

\subsection{Radiative accelerations} \label{meth-radacc}

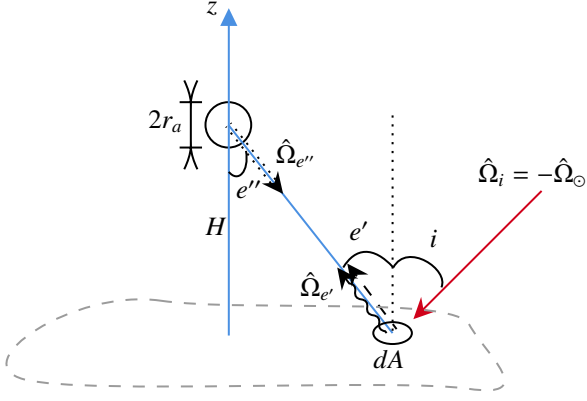
\begin{figure}
  \centering
  \tikzset{every picture/.style={line width=0.75pt}} %set default line width to 0.75pt        
  
  \begin{tikzpicture}[x=0.75pt,y=0.75pt,yscale=-1,xscale=1]
  
  \draw    (139.15, 85.25) circle [x radius= 11.41, y radius= 11.41]  ;
  \draw [color={rgb, 255:red, 74; green, 144; blue, 226 }  ,draw opacity=1 ]   (139.38,191.22) -- (139.38,28.56) ;
  \draw [shift={(139.38,26.56)}, rotate = 450] [fill={rgb, 255:red, 74; green, 144; blue, 226 }  ,fill opacity=1 ][line width=0.75]  [draw opacity=0] (8.93,-4.29) -- (0,0) -- (8.93,4.29) -- cycle    ;
  
  \draw [color={rgb, 255:red, 74; green, 144; blue, 226 }  ,draw opacity=1 ]   (139.15,85.25) -- (221.13,189.99) ;

  \draw    (120.42,73.84) -- (120.42,96.66) ;
  \draw [shift={(120.42,96.66)}, rotate = 90] [color={rgb, 255:red, 0; green, 0; blue, 0 }  ][line width=0.75]    (0,5.59) -- (0,-5.59)(10.93,-3.29) .. controls (6.95,-1.4) and (3.31,-0.3) .. (0,0) .. controls (3.31,0.3) and (6.95,1.4) .. (10.93,3.29)   ;
  \draw [shift={(120.42,73.84)}, rotate = 270] [color={rgb, 255:red, 0; green, 0; blue, 0 }  ][line width=0.75]    (0,5.59) -- (0,-5.59)(10.93,-3.29) .. controls (6.95,-1.4) and (3.31,-0.3) .. (0,0) .. controls (3.31,0.3) and (6.95,1.4) .. (10.93,3.29)   ;
  \draw    (221.13, 189.99) circle [x radius= 9.63, y radius= 5.5]  ;
  \draw [color={rgb, 255:red, 208; green, 2; blue, 27 }  ,draw opacity=1 ]   (232.91,181.58) -- (295.52,118.49) ;
  
  \draw [shift={(231.5,183)}, rotate = 314.78] [fill={rgb, 255:red, 208; green, 2; blue, 27 }  ,fill opacity=1 ][line width=0.75]  [draw opacity=0] (10.72,-5.15) -- (0,0) -- (10.72,5.15) -- (7.12,0) -- cycle    ;
  \draw  [dash pattern={on 0.84pt off 2.51pt}]  (221.13,80.5) -- (221.13,189.99) ;

  \draw    (221.5,157) .. controls (228.5,144) and (249.5,158) .. (245.5,167) ;

  \draw    (221.5,157) .. controls (211.5,146) and (202.5,147) .. (196.5,157) ;

  \draw [color={rgb, 255:red, 0; green, 0; blue, 0 }  ,draw opacity=1 ] [dash pattern={on 4.5pt off 4.5pt}]  (199.7,156.6) -- (223.13,187.99) ;
  
  \draw [shift={(198.5,155)}, rotate = 53.26] [fill={rgb, 255:red, 0; green, 0; blue, 0 }  ,fill opacity=1 ][line width=0.75]  [draw opacity=0] (10.72,-5.15) -- (0,0) -- (10.72,5.15) -- (7.12,0) -- cycle    ;
  \draw [color={rgb, 255:red, 0; green, 0; blue, 0 }  ,draw opacity=1 ] [dash pattern={on 0.84pt off 2.51pt}]  (160.37,114.64) -- (137.97,86.18)(162.73,112.79) -- (140.33,84.32) ;
  
  \draw [shift={(166.5,120)}, rotate = 231.8] [fill={rgb, 255:red, 0; green, 0; blue, 0 }  ,fill opacity=1 ][line width=0.75]  [draw opacity=0] (10.72,-5.15) -- (0,0) -- (10.72,5.15) -- (7.12,0) -- cycle    ;
  \draw    (147.96,98.35) .. controls (149.68,108.89) and (143.95,114.57) .. (139.38,108.89) ;

  \draw [color={rgb, 255:red, 0; green, 0; blue, 0 }  ,draw opacity=1 ]   (194.7,158.6) -- (199.48,165.01) .. controls (201.81,165.35) and (202.81,166.69) .. (202.47,169.02) .. controls (202.13,171.35) and (203.13,172.69) .. (205.46,173.03) .. controls (207.79,173.36) and (208.79,174.7) .. (208.46,177.03) .. controls (208.12,179.36) and (209.12,180.7) .. (211.45,181.04) .. controls (213.78,181.38) and (214.78,182.72) .. (214.44,185.05) .. controls (214.1,187.38) and (215.1,188.72) .. (217.43,189.05) -- (218.13,189.99) -- (218.13,189.99) ;
  
  \draw [shift={(193.5,157)}, rotate = 53.26] [fill={rgb, 255:red, 0; green, 0; blue, 0 }  ,fill opacity=1 ][line width=0.75]  [draw opacity=0] (10.72,-5.15) -- (0,0) -- (10.72,5.15) -- (7.12,0) -- cycle    ;
  \draw  [color={rgb, 255:red, 155; green, 155; blue, 155 }  ,draw opacity=1 ][dash pattern={on 4.5pt off 4.5pt}] (105,173.65) .. controls (147.97,179.7) and (251.36,170.97) .. (259.42,185.74) .. controls (267.48,200.51) and (295.01,223.34) .. (258.75,217.97) .. controls (222.49,212.59) and (67.74,221.32) .. (37.19,213.27) .. controls (6.64,205.21) and (62.03,167.61) .. (105,173.65) -- cycle ;
  
%  \draw (190.06,123.44) node   {$R$};
  \draw (132.86,136.29) node   {$H$};
  \draw (107.84,83.57) node   {$2 r_a$};
  \draw (219.33,203.34) node   {$dA$};
  \draw (131.07,26.6) node   {$z$};
  \draw (241.59,142.58) node   {$i$};
  \draw (204.59,138.58) node   {$e'$};
  \draw (184.06,166.44) node   {$\hat{\Omega}_{e'}$};
  \draw (173.06,99.44) node   {$\hat{\Omega}_{e''}$};
  \draw (291.06,109.44) node   {$\hat{\Omega}_{i} =-\hat{\Omega}_{\odot }$};
  \draw (149.59,117.58) node   {$e''$};

  \end{tikzpicture}
  \caption{Schematic diagram of a spherical particle hovering over the surface. The particle of radius $ r_a $ hovers over the planar regolith at the height of $ H (\gg r_a $). The solar irradiation (the long red arrow) is originated from the direction $ \hat{\Omega}_i $. A tiny surface patch with an area of $ dA $ reflects the solar radiation and emits the thermal radiation to the particle (the dashed and squiggly arrows, respectively, to the direction of $ \hat{\Omega}_{e'} $) with the emission angle of $ e' $. In our model, $ e'' = e' $ holds because of the planar surface assumption. The isothermal regolith region (gray dashed) is assumed to be a circle with a radius of $ r_0 (\gg H \gg r_a) $. No radiation outside this region is considered because the patch right below the particle already fills most of the field of view ($ r_0 \gg H $).}
  \label{fig:fschem_ref_th}
\end{figure}

To derive the formalism for radiative accelerations, we consider an arbitrary spectral irradiance ($ J_\lambda $) from a black body of temperature $ T $, spectrum-averaged emissivity $ \bar{\epsilon} $, and solid angle $ d\Omega $ on the particle from a certain incident direction (Fig \ref{fig:fschem_ref_th}). For simplicity, we define a few auxiliary quantities (Appendix \ref{app: acc notation}):
\begin{equation} \label{eq: bar Qpr}
  \bar{Q}_\mathrm{pr} (r_a, T, \epsilon) 
    \equiv \int_{0}^{\infty} Q_\mathrm{pr}(\lambda, r_a) b_\lambda(\lambda, T, \epsilon) d\lambda ~,
\end{equation}
\begin{equation}\label{eq: b lambda}
  b_\lambda (\lambda, T, \epsilon) 
    = \frac{\pi}{\sigma_\mathrm{SB} T^4} \frac{ \epsilon(\lambda)}{\bar{\epsilon}(T, \epsilon)} B_\lambda(\lambda, T) ~,
\end{equation}
\begin{equation} \label{eq: epsilon-bar}
  \bar{\epsilon} (T, \epsilon)
    = \frac{\pi}{\sigma_\mathrm{SB} T^4} \int_{0}^{\infty} \epsilon(\lambda) B_\lambda(\lambda, T) d\lambda ~.
\end{equation}
Here, $ B_\lambda(\lambda, T) $ is the Planck function, $ \sigma_\mathrm{SB} $ is the Stefan--Boltzmann constant, and $ \epsilon(\lambda) $ is the emissivity function of the radiating source (either the Sun or the AR). Then, $ \bar{Q}_\mathrm{pr} $ in Eq. (\ref{eq: bar Qpr}) is the spectrum-averaged $ Q_\mathrm{pr} $ value, $ b_\lambda $ in Eq. (\ref{eq: b lambda}) is the normalized Planck function, and $ \bar{\epsilon} $ is the spectrum-averaged emissivity. Using these parameters, the magnitude of the radiative acceleration on a Mie particle of radius $ r_a $ and mass density $ \rho_m $ is (Appendix \ref{app: acc acc}):
\begin{equation}\label{eq: a_rad}
  a = \tilde{C} \frac{\bar{\epsilon} \bar{Q}_\mathrm{pr}(r_a, T)}{r_a \rho_m} T^4 \frac{d \Omega}{\pi} ~.
\end{equation}
%See Appendix \ref{app: acc acc} and \ref{app: acc notation} for detailed description about the auxiliary quantities and the derivation of Eq. (\ref{eq: a_rad}).
%In the practical calculation, $ \bar{\epsilon} = \epsilon = 0.90 $ is fixed, and $ \bar{Q}_\mathrm{pr} $ is calculated in advance for various points in the $ (r_a, T) $ space (see below) to replace the integral in the left hand side.

\begin{figure}
  \centering
%  \sidecaption
  \includegraphics[width=\linewidth]{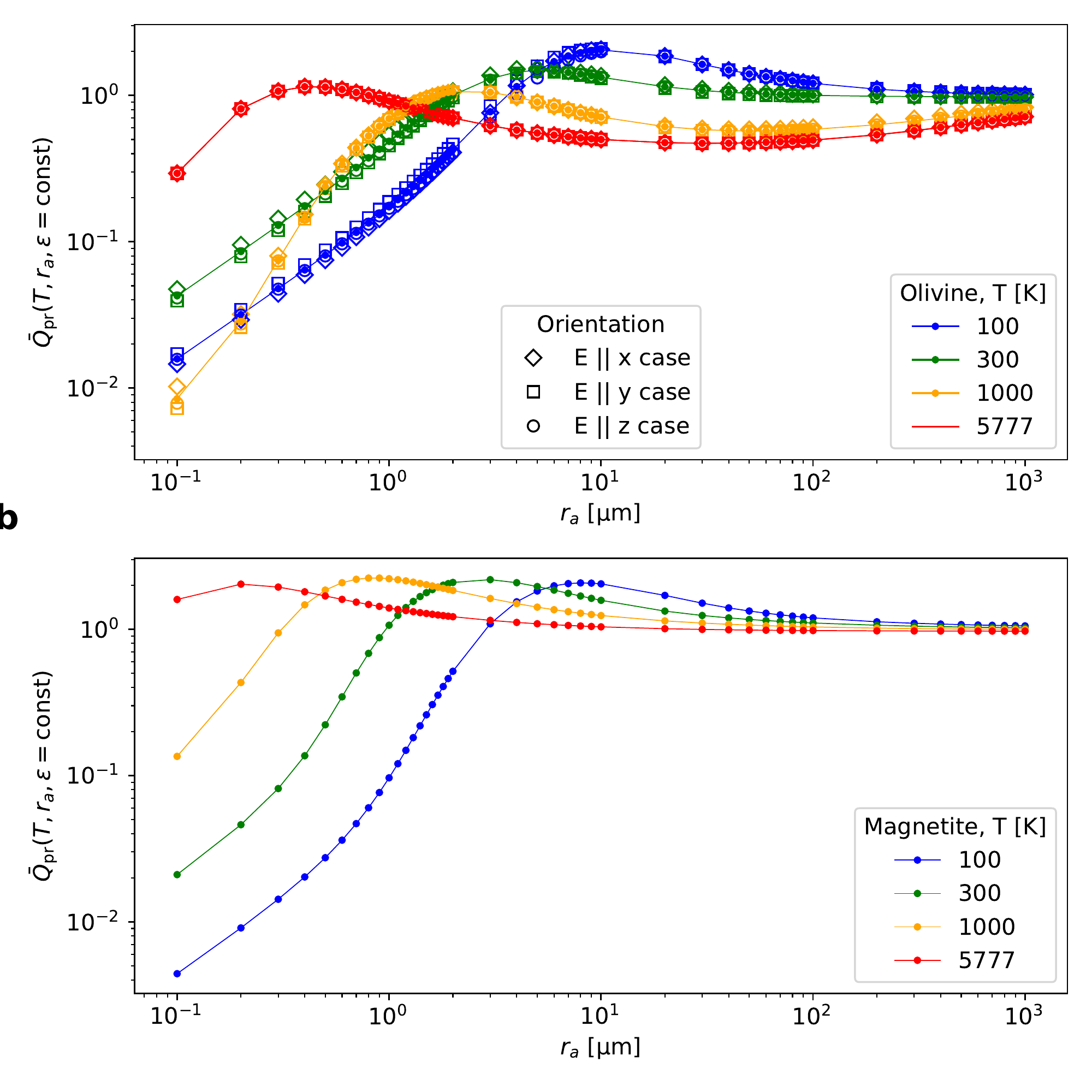}
  \caption{Calculated $ Q_\mathrm{pr} $ value averaged over blackbody spectra of different temperatures ($ \bar{Q}_\mathrm{pr} $ in our notation). The $ \bar{Q}_\mathrm{pr} $ values are plotted as a function of the particle size $ r_a $ and blackbody temperature $ T $ of the incident radiation for two minerals: \textbf{a}, olivine and \textbf{b}, magnetite. The emissivity $ \epsilon $ is assumed to be constant, so $ \bar{\epsilon} = \epsilon $ cancels out in Eq. (\ref{eq: b lambda}). For olivine, three polarization directions with respect to the crystalline axes can lead to small differences in the results (three different markers; see Appendix \ref{app: qpr}), and the average value (solid line) is used in this work. $ \bar{Q}_\mathrm{pr} $ converges to unity for a large particle limit, as predicted from the geometrical optics approximation.}
  \label{fig:qprbar}
\end{figure}

We use the $ Q_\mathrm{pr} $ values calculated in Appendix \ref{app: qpr} to integrate in Eq. (\ref{eq: bar Qpr}). Figure \ref{fig:qprbar} shows the calculated $ \bar{Q}_\mathrm{pr} $ for magnetite and olivine when $ \epsilon $, and therefore $ \bar{\epsilon} $, are constants. The trend that $ \bar{Q}_\mathrm{pr} \rightarrow 1 $ as $ r_a \gg \lambda $ (i.e., the geometric optics regime) is clearly seen in Fig. \ref{fig:qprbar}. For olivine, the optical index is measured separately for three polarization directions with respect to the crystalline axis orientations. (\citealt{2001A&A...378..228F}; see Appendix \ref{app: qpr} for details). The calculated $ \bar{Q}_\mathrm{pr} $ values in Fig. \ref{fig:qprbar}\textbf{a} are, however, nearly the same for all three cases, at least under the temperatures in which we are interested (AR temperature of $ T_S \sim 100 \mathrm{-} 1000\,\mathrm{K} $ and $ T = T_\odot = 5777 \,\mathrm{K} $). Thus, the average of the three values is adopted throughout this work. Magnetite does not have multiple optical indices, because of its simple symmetric (face-centered cubic) structure.

The $ \bar{Q}_\mathrm{pr} $ values are calculated in the range of $ r_a \in [0.1, 1000] \um $ (at $ 0.1 \um $ intervals from 0.1 to $ 2.0 \um $, $ 1 \um $ intervals to $ 10 \um $, $ 10 \um $ intervals to $ 100 \um $, and $ 100 \um $ intervals to $ 1000 \um $) and $ T_S \in [10, 1300]\,\mathrm{K} $ (with $ 1\,\mathrm{K} $ interval) for both magnetite and olivine. The $ \bar{Q}_\mathrm{pr} $ values are then linearly interpolated when necessary. Figure \ref{fig:qprbar} shows $ \bar{Q}_\mathrm{pr} $ values with respect to $ r_a $ for fixed $ T $. %By doing so, one can calculate the actual magnitude of acceleration terms at each longitude and latitude on the asteroid.

\subsubsection{Solar radiation $ \a_\odot $} \label{meth-a_odot}
%-------------------------------------------------------------------------------
The irradiance from the Sun is directed inward to the AR surface with an incident angle $ i $. If it is assumed to be a uniform beam from the solid angle of $ \Delta \Omega_\odot (\rh) = \pi R_\odot^2 / \rh^2 $, Eq (\ref{eq: a_rad}) gives
\begin{equation} \label{eq: acc_odot}
\begin{aligned}
  a_\odot 
    &= \tilde{C} R_\odot^2 T_\odot^4 \frac{\bar{\epsilon}_\odot \bar{Q}_\mathrm{pr, \odot}(r_a)}{r_a \rho_m \rh^2}  \\
    &= 1.13903 \times 
      \frac{\bar{\epsilon}_\odot \bar{Q}_\mathrm{pr, \odot}(r_a)}
      {
      \left ( \frac{r_a}{1 \um} \right ) 
      \left ( \frac{\rho_m}{3000 \den} \right )
      \left ( \frac{\rh}{1 \au} \right )^{2} 
      }
     \,\mathrm{[mm/s^2]} ~.
\end{aligned}
\end{equation}
%where $ a_\odot = \| \a_{\odot} \left (r_a, \rh \,\middle |\, \mathcal{I}_a) \right \| $. $ \mathcal{I}_a $ stands for information about physical properties of the particle (refractive index). 
The solar radiation is assumed as a perfect blackbody in this work, so $ \bar{\epsilon}_\odot = \epsilon = 1 $. $ a_\odot $ decreases as $ \rh^{-2} $ as the solar irradiance decreases. For the Sun, we use the notation $ \bar{Q}_\mathrm{pr, \odot}(r_a) := \bar{Q}_\mathrm{pr}(r_a, T = T_\odot) $. 

Because $ \a_\odot $ is oriented in the anti-solar direction, $ \a_\odot $ is in the inward direction on the dayside of the asteroid. When the particle goes behind the shadow of the asteroid, we regard $ a_\odot = 0 $.

\subsubsection{Reflected solar radiation $ \a_\mathrm{ref} $} \label{meth-a_ref}
%-------------------------------------------------------------------------------
For a reflected (scattered) component of the acceleration, the radiative acceleration contributed from each infinitesimal surface patch in the field of view must be integrated. After proper integration (Appendix \ref{app: acc_ref}), the acceleration from a Lambertian scattering AR is derived as
\begin{equation} \label{eq: acc_ref}
\begin{aligned}
  a_{\mathrm{ref}} 
    &= A_B \mu_i \tilde{H} \left (\frac{H}{r_0} \right )  a_\odot(r_a, \rh) ~,
\end{aligned}
\end{equation}
where $ \mu_i := \mathrm{max} \, \{\cos i,\, 0\} $ for the incidence angle $ i $, %$ a_{\mathrm{ref}} = \left \| \a_{\mathrm{ref}}(r_a, \rh, A_B, H/r_0 \,\middle |\, \mathcal{I}_a) \right \| $, 
and the height factor
\begin{equation} \label{eq: H factor}
  \tilde{H} \left (\frac{H}{r_0} \right ) := \frac{1}{1 + (H / r_0)^2} ~.
\end{equation}
The $ r_0 $ is set as $ 1\,\% $ of the asteroid's radius throughout this work. Here, $ \tilde{H} < 1 $ but $ \tilde{H}$ is almost unity at the initial condition, as $ r_0 \gg H = 1 \cm $. We did not assume $ \tilde{H} = 1 $ in this general formula, because it plays an important role in the trajectory integration (Sect. \ref{disc-trajectory}) when $ H \gtrsim r_0 $. Because $ A_B \ll 1 $ for many asteroids, one obtains $ a_\mathrm{ref} \ll a_\odot $ as expected. 
%The direction of $ \a_\mathrm{ref} $ is always positive along the radial direction unless it is 0.
%The $ r_0 $ is selected as $ 1\,\% $ of the asteroid's radius (see discussion in Appendix \ref{app-r_0}).

\subsubsection{Thermal radiation $ \a_\mathrm{ther} $}\label{meth-a_ther}
%-------------------------------------------------------------------------------
Similar to $ a_\mathrm{ref} $, an integration must take place over the surface patches in the field of view (Appendix \ref{app: acc_ther}):
\begin{equation}\label{eq: acc_ther}
\begin{aligned}
  a_\mathrm{ther} 
    &= \tilde{C} \tilde{H} \left ( \frac{H}{r_0} \right ) \frac{\bar{\epsilon}_S \bar{Q}_\mathrm{pr} T_S^4}{r_a \rho_m} \\
    &= 2.95536 \times 
      \frac{\tilde{H} \bar{\epsilon}_S \bar{Q}_\mathrm{pr} (r_a, T_S) }
        {\left ( \frac{r_a}{1 \um} \right ) 
         \left ( \frac{\rho_m}{3000 \den} \right )}
      \left ( \frac{T_S}{500\,\mathrm{K}} \right )^4
      \,\mathrm{[mm/s^2]} ~.
\end{aligned}
\end{equation}
%where $ a_\mathrm{ther} = \| \a_\mathrm{ther} \left ( r_a, T_S, H/r_0 \, \middle |\, \mathcal{I}_a, \mathcal{I}_\mathrm{AR} \right ) \| $. $ \mathcal{I}_\mathrm{AR} $ stands for the information on AR, such as the thermal parameter $ \Theta $. 
From the formula, we see that $ a_\mathrm{ther} $ is very sensitive to the AR temperature $ T_S $. Also, it is linearly proportional to $ \bar{Q}_\mathrm{pr} $, which implies the importance of the adequate Mie calculation, especially for small particles (when $ \bar{Q}_\mathrm{pr} $ deviates significantly from unity).

\subsubsection{Comparing $ \a_\odot $ and $ \a_\mathrm{ther} $} \label{meth-a_ther/a_odot z}
%%%%%%%%%%%%%%%%%%%%%%%%%%%%%%%%%%%%%%%%%%%%%%%%%%%%%%%%%%%%%%%%%%%%%%%%%%%%%%%%
To compare the $ z $-axis components of $ \a_\odot $ in Eq. (\ref{eq: acc_odot}) and $ \a_\mathrm{ther} $ in Eq. (\ref{eq: acc_ther}), we re-formulated $ a_\mathrm{ther} $ using Eqs. (\ref{eq: T_eqm}) and (\ref{eq: acc_odot}):
\begin{equation}
  a_\mathrm{ther} = a_\odot \tilde{H} \left (\frac{H}{r_0} \right ) R_Q(r_a, T_S) \frac{1 - A_B}{\bar{\epsilon}_\odot} \left ( \frac{T_S}{T_\mathrm{eqm}} \right )^4 ~,
\end{equation}
where $ R_Q(r_a, T_S) := \bar{Q}_\mathrm{pr}(r_a, T_S) / \bar{Q}_\mathrm{pr, \odot}(r_a) $ is the \textit{relative efficacy} of the thermal radiation from the AR compared to the solar radiation. In other words, it can be understood as the ratio of \textit{effective} cross-sectional areas of the particle exposed to the spectrum of temperature $ T_S $ and $ T_\odot $. 
%It is plotted in Fig. \ref{fig:RQplot} for olivine and magnetite for some selected AR temperatures ($ T_S $). 
The ratio of the two acceleration mechanisms along the $ z $-axis is then given by
\begin{equation}\label{eq: a_ther/a_odot z}
  \frac{a_\mathrm{ther}}{a_{\odot, z}} 
    =\tilde{H} \left (\frac{H}{r_0} \right )
      R_Q(r_a, T_S)
      \frac{1-A_B}{\bar{\epsilon}_\odot} 
      \frac{1}{\mu_i}
      \left (\frac{T_S}{T_\mathrm{eqm}}\right )^4
      ~.
\end{equation} 
Here, $ a_{\odot, z} = a_\odot \mu_i $. 

If the thermal inertia $ \Gamma = 0 $, the surface temperature is the same as the equilibrium temperature, $ T_S = T_\mathrm{eqm} $. This is the case for the near-Earth asteroid thermal model (NEATM; \citealt{1998Icar..131..291H}) for example, with the beaming parameter of unity. Thus, the ratio in Eq. (\ref{eq: a_ther/a_odot z}) stays constant during rotation of the asteroid  for $ \Gamma = 0 $ unless $ \mu_i $ vanishes. In reality, however, this ratio changes over the local time. In particular, it increases towards the evening terminator. This is because $ T_S $ does not cool immediately like NEATM (Sect. \ref{res-temp prof}), and so $ T_S > T_\mathrm{eqm} \mu_i^{1/4} $. Therefore, one would expect $ a_\mathrm{ther} $ to dominate $ a_\odot $ in the afternoon under TPM, which is a more realistic version of NEATM. This is a key point in the motivation of the present study.

\subsection{Gravity and rotation} \label{meth-grav and cen}
The gravitational acceleration is calculated by
\begin{equation}\label{eq: a_grav}
\begin{aligned}
  a_\mathrm{grav} &= \frac{2 \pi G}{3} \rho_M D \\
    &= 0.27957 \times 
      \left ( \frac{D}{1 \km} \right ) 
      \left ( \frac{\rho_M}{2000 \den} \right ) 
      \,\mathrm{[mm/s^2]} ~,
\end{aligned}
\end{equation}
where $ \rho_M $ is the bulk mass density of the asteroid (not necessarily equal to the mass density of grains $ \rho_m $; likely $ \rho_M < \rho_m $ due to the bulk porosity), and $ D $ is the asteroid diameter. The gravitational acceleration is directed inward from the AR. Figure 5 of \cite{2012AJ....143...66J} can be reproduced by comparing $ a_\mathrm{grav} $ in Eq. (\ref{eq: a_grav}) with $ a_\odot $ in Eq. (\ref{eq: acc_odot}).

The last remaining acceleration component is a fictitious one, namely the centrifugal acceleration. This is introduced only when describing the net acceleration of the particle from an observer sitting on the AR (i.e.,  in the rotating frame on the asteroid). The centrifugal acceleration on the equator is calculated from
\begin{equation}\label{eq: a_cen}
\begin{aligned}
  a_\mathrm{cen} &= \frac{2 \pi^2 D}{P_\mathrm{rot}^2}\\
    &= 1.52309 \times 
      \left ( \frac{D}{1 \km} \right )
      \left ( \frac{P_\mathrm{rot}}{1 \h} \right )^{-2}
      \,\mathrm{[mm/s^2]} ~,
\end{aligned}
\end{equation}
for the asteroid's rotational period $ P_\mathrm{rot} $. In the main part of this work, $ a_\mathrm{cen} $ is taken into account for calculating the \textit{effective} gravity by $ (\a_\mathrm{grav} + \a_\mathrm{cen}) \cdot \rhat = a_\mathrm{grav} - a_\mathrm{cen} $. As we are considering an initially stationary particle with respect to the observer, the Coriolis force is zero.

\subsection{Parameter selection of model asteroids} \label{meth-model param selection}
%%%%%%%%%%%%%%%%%%%%%%%%%%%%%%%%%%%%%%%%%%%%%%%%%%%%%%%%%%%%%%%%%%%%%%%%%%%%%%%%
For fictitious model asteroids, two aspect angles (the angle between the spin vector and the asteroid--Sun vector) are considered, namely, $ 90^\circ $ and $ 45^\circ $, which are called the ``perpendicular'' and ``aspect $ 45^\circ $'' models in this work, respectively. It is assumed that the fiducial model asteroids are located at a heliocentric distance of $ \rh = 0.2 \au $, that is, in the near-Sun environment. We further changed it from $ 0.1 $ to $ 2 \au $ for test cases (25 equally spaced $ \rh $ values in logarithmic scale by $ \rh = 10^{k (\log_{10} 20)/24 - 1 } \au $ for integer $ k = 0 $ to $ 24 $). Fiducial asteroids have smooth spherical shapes with $ D = 1 \km $, and for test cases, $ D $ ranges from 0.1 to 80 km (50 equally spaced $ D $ values in logarithmic scale by $ D = 10^{k (\log_{10} 800)/49 - 1 } \km $ for integer $ k = 0 $ to $ 49 $). For S, B, and C-type asteroids, the bulk mass density $ \rho_M \sim 2660 \pm 1290 $, $2190 \pm 1000,$ and $ 1570 \pm 1380 \den$, respectively \citep{2012P&SS...73...98C}. Therefore, as a simplifying assumption, the model asteroids have a fixed $ \rho_M = 2000 \den $. The Bond albedo is fixed at $ A_B = 0.05 $, a typical value for an asteroid with $ p_\mathrm{V} \sim 0.1\mathrm{-}0.2 $ using Eq. (\ref{eq: A_B}). The rotation period of the fiducial model is also fixed to $ P_\mathrm{rot} = 6 \h $, a representative value of the asteroids in the size range \citep[][updated in 2020\footnote{\url{http://www.minorplanet.info/lightcurvedatabase.html}}]{2009Icar..202..134W}; the $ P_\mathrm{rot} $ value has almost no effect on these preliminary results when it is changed by a factor of two (see Sect. \ref{res-size-size}). The thermal inertia is taken as $ 200\tiu $ (a typical value for asteroids of a similar size range in \citealt{2015aste.book..107D}). 

All the above-mentioned parameters are summarized in Table \ref{tab: phys_phae} (Model asteroids columns). The surface temperatures of these asteroids are calculated by solving the 1D heat conduction equation using these parameters (Sect. \ref{meth-tpm}).

\begin{table*}[h!]
\caption{Physical parameters of Phaethon \citep{2018A&A...620L...8H} and the fiducial and tested parameters for fictitious model asteroids are given. 
%The perpendicular model is set to have arbitrary geometric albedo and slope parameter so that the Bond albedo is 0.10. 
}
\centering
  \begin{tabular}{lccccc}
        \hline
        Parameter                          &           Symbol           &           Phaethon           &       \multicolumn{2}{c}{Model asteroids}        &                       Unit                       \\
                                           &                            &                              &       fiducial       &       tested values       &                                                  \\ \hline
        True anomaly                       &           $ f $            &     $ 0 \mathrm{-} 360 $     &          -           &             -             &                    $ ^\circ $                    \\
        Heliocentric distance              &          $ \rh $           &   $ 0.14 \mathrm{-} 2.40 $   &       $ 0.2 $        &   $ 0.1\mathrm{-}2 $    &                        au                        \\ \hline
        Diameter                           &           $ D $            &             5.1              &        $ 1 $         &    $ 0.1\mathrm{-}80 $    &                        km                        \\
        Geometric albedo                   &      $ p_\mathrm{V} $      &             0.12             &          -           &             -             &                        -                         \\
        Slope parameter                    &           $ G $            &             0.15             &          -           &             -             &                        -                         \\
        Bond albedo                        &          $ A_B $           &            0.046             &         0.05         &           0.05            &                        -                         \\
        Bulk mass density                  &         $ \rho_M $         &            1,670             &        2,000         &           2,000           &               $ \mathrm{kg/m^3} $                \\
        Spin axis (ecliptic)               & $ (\lambda_s,\, \beta_s) $ & $ (318^\circ,\, -47^\circ) $ & \multicolumn{2}{c}{aspect angle $ 90^\circ $/$ 45^\circ $} &                    $ ^\circ $                    \\
        %    Aspect angle
%                &  $ \theta_\mathrm{asp} $   &             101              &      $ ^\circ $      &  \\
        Rotational period                  &     $ P_\mathrm{rot} $     &           3.603957           &          6           &           3, 6            &                       hour                       \\
        Thermal inertia                    &         $ \Gamma $         &             600              &         200          &            200            & $ \mathrm{tiu} \equiv \mathrm{J/m^2/K/s^{1/2}} $ \\
        Thermal parameter\tablefootmark{a} &         $ \Theta $         &   $ 0.21 \mathrm{-} 15.1 $   &      $ 0.094 $       & $ 0.033 \mathrm{-} 1.27 $ & $ \mathrm{tiu} \equiv \mathrm{J/m^2/K/s^{1/2}} $ \\ \hline
        %    Composition
%                 &             -              &       \multicolumn{2}{c}{olivine, magnetite}        &             -             &  \\
        %    Mass density
%                &          $\rho_m$          &              \multicolumn{2}{c}{3,000}              &    $ \mathrm{kg/m^3} $    &  \\
        %    Radius
%                      &           $r_a$            &      \multicolumn{2}{c}{$ 0.5 \mathrm{-} 40 $}      &    $ \um $     &  \\
        %    Initial co-latitude
%         &          $\theta$          &    $ 60 \mathrm{-} 120 $     &      $ ^\circ $      &  \\
        %    Initial longitude
%           &           $\phi$           &     $ 0 \mathrm{-} 360 $     & $ 0 \mathrm{-} 360 $ &        $ ^\circ $         &  \\
        %    Initial height
%              &           $ H $            &             0.01             &         0.01         &      $ \mathrm{m} $       &  \\ \hline
  \end{tabular}
  \tablefoot{\tablefoottext{a}{The thermal parameter is calculated using Eq. (\ref{eq: theta-def}) with $ \bar{\epsilon}_S = 0.90 $.}}
  \label{tab: phys_phae}
\end{table*}

\subsection{Selection of particle parameters } \label{meth-particle param selection}
%%%%%%%%%%%%%%%%%%%%%%%%%%%%%%%%%%%%%%%%%%%%%%%%%%%%%%%%%%%%%%%%%%%%%%%%%%%%%%%%
The particles have mass densities of $ \rho_m = 3000 \den $ following previous works on dust ejection or electrostatic dust lofting simulations \citep[e.g.,][]{2013AJ....145..154L, 2013ApJ...771L..36J, 2015P&SS..116...18S, 2017AJ....153...23H, 2018AdSpR..62..896O}. The assumed mass density $ \rho_m $ is close to the (25143) Itokawa sample ($ 3400 \den $ \citealt{2011Sci...333.1125T}). The particles are assumed to be homogeneous\footnote{
  The compact material density (the density when microporosity is zero) is $ \sim 3,200 \den $ for forsterite (non-Fe olivine; $\mathrm{Mg_{2} Si O_4}$), $ \sim 4,400 \den $ for fayalite (non-Mg olivine; $\mathrm{Fe_{2} Si O_4}$), and $ \sim 5,200 \den $ for iron (II, III) oxide (pure $ \mathrm{Fe_3 O_4} $; all densities from \cite{robie1962}). The adopted density $ \rho_m = 3000 \den $ can therefore  be understood as a {homogeneously porous sphere} with certain microporosity in the Mie theory.
}
and spherical, and placed on the surface (at a height of $ H = 1\cm $ over the surface). The radiation pressure coefficient ($ Q_\mathrm{pr} $) is obtained from Mie theory (Sect. \ref{meth-mie}).

We do not consider microporosity ($ \phi $) in this work. If $ \phi $ is to be introduced, the particle density $ \rho_m $ must be updated to $ (1 - \phi)\rho_m $, but the Mie theory (Sect. \ref{meth-mie}) still applies if the particle is still assumed to be {homogeneous}. The radiative accelerations are then affected linearly: $ a_\odot $, $ a_\mathrm{ref} $ and $ a_\mathrm{ther} $ (Eqs. \ref{eq: acc_odot}, \ref{eq: acc_ref}, \ref{eq: acc_ther}, respectively) have the simple identical $ \rho_m^{-1}  $ dependency, and hence the net {radiative} acceleration will be scaled as $ a_\mathrm{rad} \propto \rho_m^{-1} \propto (1 - \phi)^{-1}$.

%%%%%%%%%%%%%%%%%%%%%%%%%%%%%%%%%%%%%%%%%%%%%%%%%%%%%%%%%%%%%%%%%%%%%%%%%%%%%%%%
\section{Results}
%%%%%%%%%%%%%%%%%%%%%%%%%%%%%%%%%%%%%%%%%%%%%%%%%%%%%%%%%%%%%%%%%%%%%%%%%%%%%%%%

\subsection{Temperature profile on asteroids} \label{res-temp prof}
\begin{figure*}[t!]
  \centering
  \includegraphics[width=\linewidth]{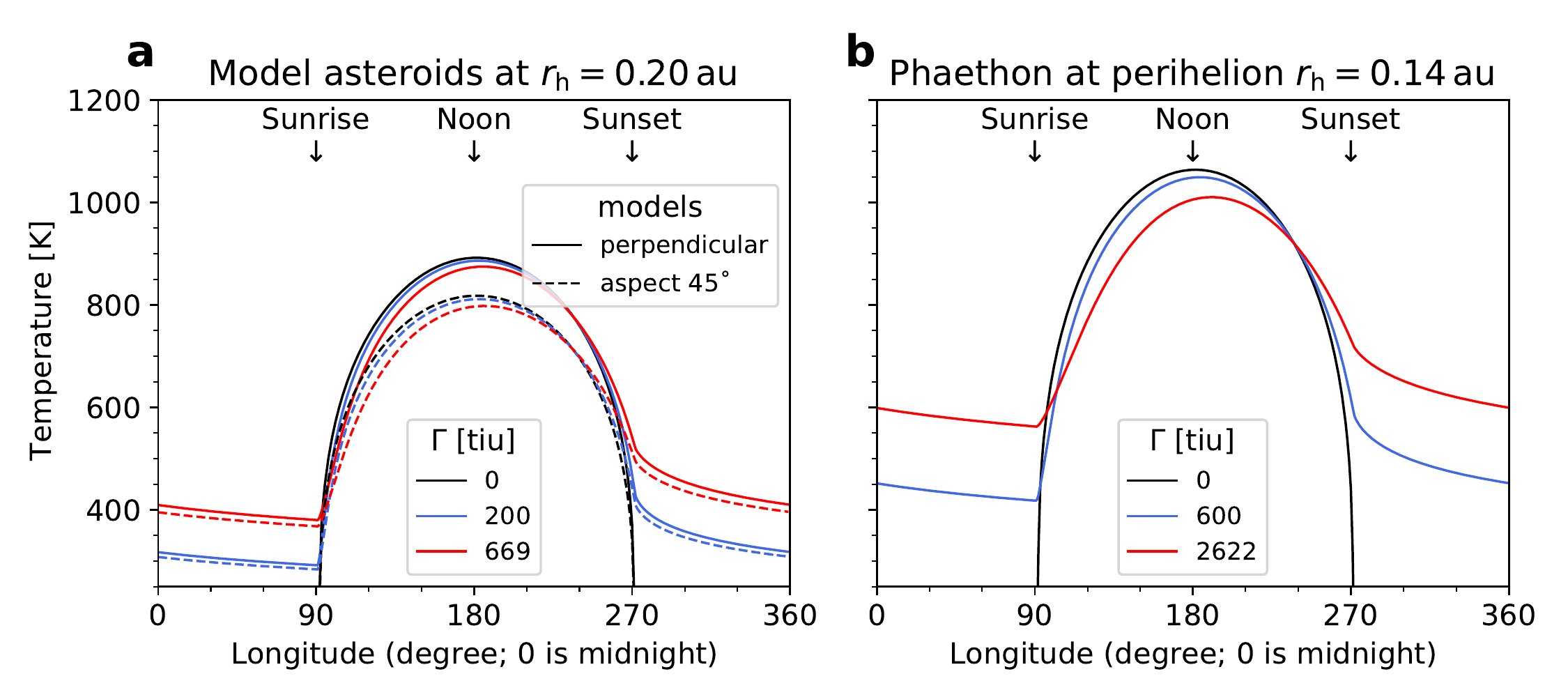}
  \caption{Equatorial temperature profiles. \textbf{a}, the fictitious asteroid at $ \rh = 0.20 \au $ and \textbf{b}, Phaethon at perihelion ($ \rh = 0.14 \au $). For the model asteroids, different aspect angles are considered: perpendicular (solid) and $ 45^\circ $ (dashed). The different colors of the lines indicate different thermal inertia values: black lines for $ \Gamma = 0 $, light \textcolor{blue}{blue} lines for fiducial value for model asteroids and Phaethon (Table \ref{tab: phys_phae}), and \textcolor{red}{red} lines for the values considering the temperature-dependent thermal inertia ($ \Gamma = 669\tiu $ and $ 2622\tiu $ for model asteroids and Phaethon, respectively; see Sect. \ref{disc-effect of TI}). }
  \label{fig:tempprof}
\end{figure*}

%%%%%%%%%%%%%%%%%%%%%%%%%%%%%%%%%%%%%%%%%%%%%%%%%%%%%%%%%%%%%%%%%%%%%%%%%%%%%%%%
To test the validity of our TPM calculation, a few temperature profiles over time were generated and compared with previous works \citep[e.g.,][Figure 2.2]{MuellerM2007PhDT}, and the code passed the tests. Temperature profiles in Fig. \ref{fig:tempprof} resemble some representative ones from fictitious model asteroids and Phaethon used in this work. The thermal inertia difference does not induce large temperature profile changes during the day compared to the night (for the physical reasons, see Sect. \ref{disc-effect of TI}).

The nighttime temperature increases significantly as thermal inertia. Also, the nighttime temperatures for perpendicular and aspect $ 45^\circ $ models are very similar, although they differ considerably in the daytime. Therefore, it is expected that $ a_\mathrm{ther} \propto T_S^4 $ would be insensitive to the aspect angle at night. The largest thermal inertias are the fiducial values corrected by the temperature dependency (Sect. \ref{disc-effect of TI}).

\subsection{Components of acceleration} \label{res-acc decompose}
\begin{figure*} [ht!]
  \centering
  \includegraphics[width=\linewidth]{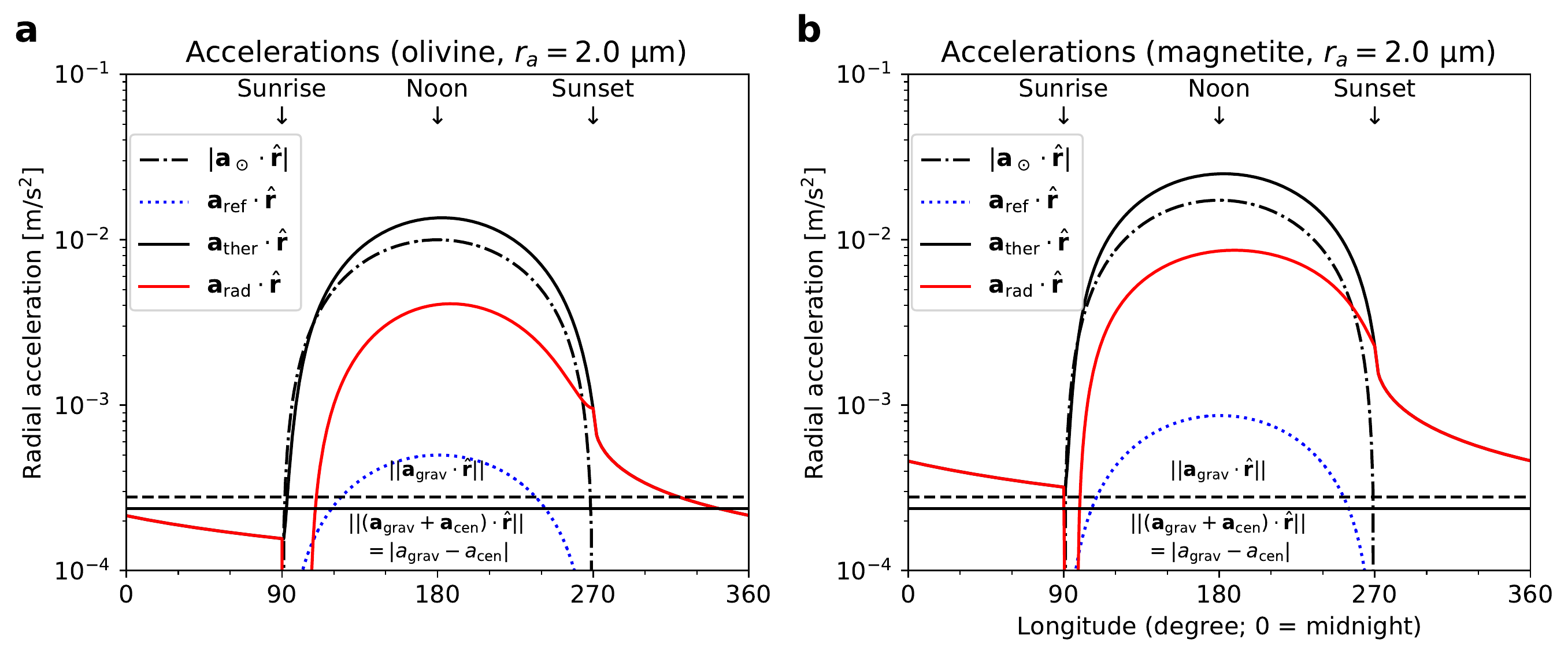}
  \caption{Components of the \textit{radiative} accelerations $ \a_\mathrm{rad} \cdot \rhat $ acting on a $ r_a = 2 \um $ particle at the height of $ H = 1\cm $ on the equator of $ D = 1 \km $ model asteroid as a function of longitude (local time), for \textbf{a}, olivine; \textbf{b}, magnetite. The black horizontal lines show the gravity (dashed) and the effective gravity taking the centrifugal acceleration into account (solid). The absolute value of $ \a_\odot \cdot \rhat $ is used as it is negative. Both $ \a_\odot $ and $ \a_\mathrm{ref} $ turn off almost immediately after Sunset, which causes the spiky feature of $ \a_\mathrm{rad}\cdot \rhat $ at longitude of $ 270^\circ $. $ \a_\mathrm{ther} $ persists throughout the night thanks to nonzero thermal inertia.}
  \label{fig:perpaccdiagramoliv2umdecompose}
\end{figure*}

%%%%%%%%%%%%%%%%%%%%%%%%%%%%%%%%%%%%%%%%%%%%%%%%%%%%%%%%%%%%%%%%%%%%%%%%%%%%%%%%
The resultant radiative acceleration components for $ r_a=2 \um $ olivine and magnetite particles on the equator of the fiducial model asteroid ($ D = 1 \km $) are shown in Fig. \ref{fig:perpaccdiagramoliv2umdecompose}. Here we see that, especially at the evening side, the thermal component, $ \a_\mathrm{ther} $, dominates the total radiative acceleration, $ \a_\mathrm{rad} $. At night, $ \a_\mathrm{rad}\cdot \rhat $ is larger than the threshold $ \| (\a_\mathrm{grav} + \a_\mathrm{cen}) \cdot \rhat \| $, that is, the particles are accelerated outwards from the AR. The difference between these curves of olivine and magnetite comes from the difference in the refractive indices (Sect. \ref{meth-mie} and Appendix \ref{app: qpr}), which makes $ \bar{Q}_\mathrm{pr} $ different (Fig. \ref{fig:qprbar}).

The sharp spike of $ \a_\mathrm{rad} \cdot \rhat $ at the evening terminator is caused by the sudden quench of the solar radiation near the terminator (when the particle enters the  shadow behind the asteroid). In the computation, $ \a_\odot $ is not turned off until it reaches this shadow even though its footprint reached the longitude of $ 270^\circ $, but this lag is invisible due to the small height $ H = 1\cm $. The reflected component $ a_\mathrm{ref} $ turns off right after Sunset, that is, when the footprint of the particle is in shadow, regardless of the particle's height.

\subsection{Acceleration over local time} \label{res-acc prof}
%%%%%%%%%%%%%%%%%%%%%%%%%%%%%%%%%%%%%%%%%%%%%%%%%%%%%%%%%%%%%%%%%%%%%%%%%%%%%%%%

\begin{figure*} [t!]
  \centering
  \includegraphics[width=\linewidth]{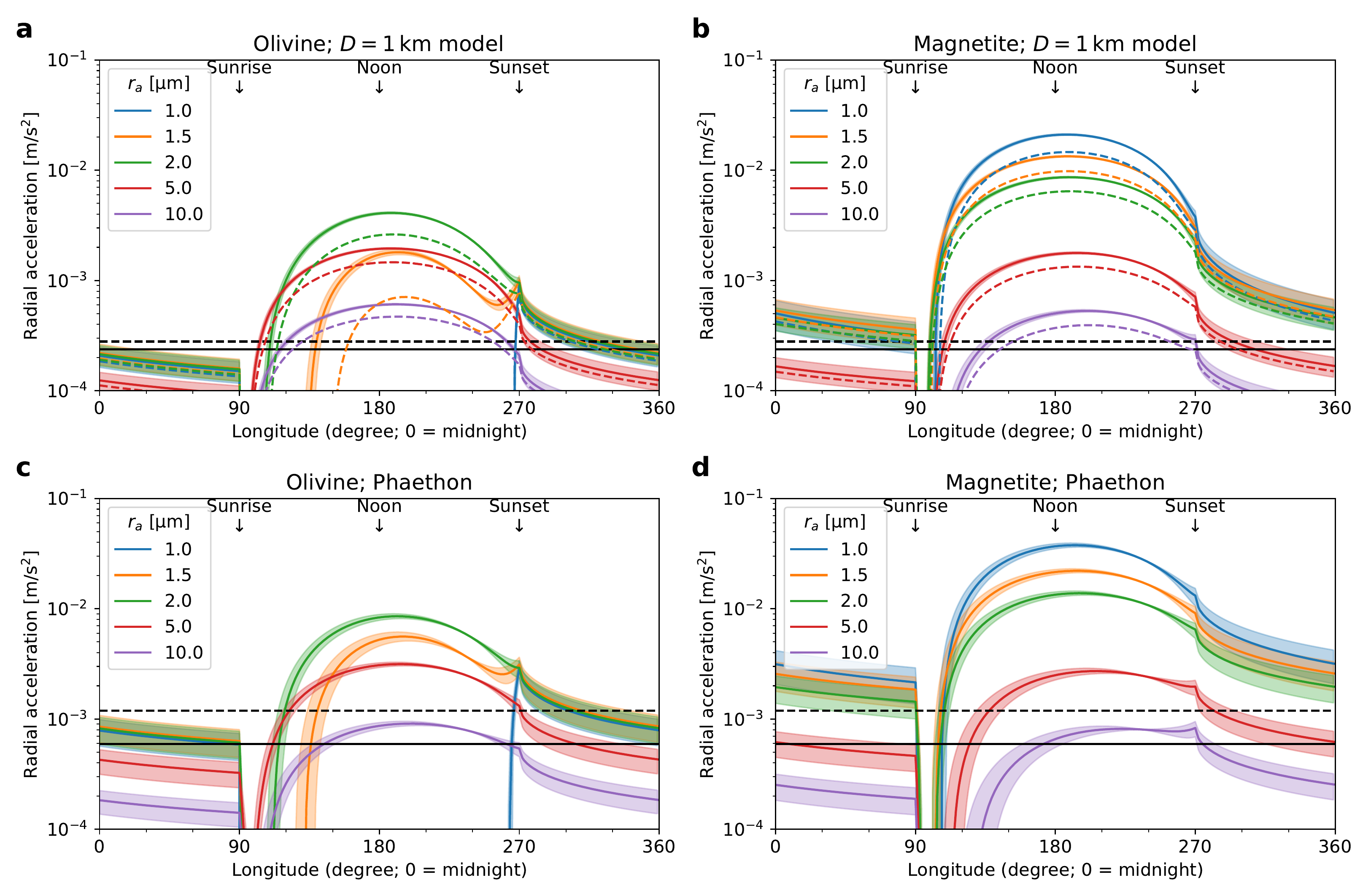}
  \caption{Radial  accelerations on the equatorial surface at heights of $ H = 1\cm $, for the corresponding particle radius $ r_a $ with respect to longitude. The net radial \textit{radiative} accelerations ($ \mathbf{a}_\mathrm{rad} \cdot \hat{\mathbf{r}} $) are shown. The black horizontal lines show the gravity and the effective gravity as in Fig. \ref{fig:perpaccdiagramoliv2umdecompose}. \textbf{a, c}, olivine; \textbf{b, d}, magnetite. Panels \textbf{a} and \textbf{b} are calculated for fiducial perpendicular (solid) and aspect $ 45^\circ $ (dashed) model asteroids ($ D = 1 \km $) at the heliocentric distance $ \rh = 0.2 \au $ for given particle radii $ r_a $ in the legend. For the perpendicular model, the $ 25 \,\% $ ambiguity in the thermal inertia ($ \Gamma = 200 \pm 50 \, \mathrm{tiu} $) is expressed as shading.
  Panels \textbf{c} and \textbf{d} are calculated for Phaethon using the nominal physical parameter values (Table \ref{tab: phys_phae}) \textit{at the perihelion}. The shading signifies the uncertainty range in the thermal inertia ($ \Gamma = 600 \pm 200 \, \mathrm{tiu} $, \citealt{2018A&A...620L...8H}) of Phaethon. 
  From these figures, it is obvious that most of $\lesssim 10 \um$-sized particles are accelerated outwards from the AR.}
  \label{fig:acc_diagram-alltifactor1}
\end{figure*}

Similar to Fig. \ref{fig:perpaccdiagramoliv2umdecompose}, the net radiative accelerations $ \a_\mathrm{rad} \cdot \rhat $ for various particle radius $ r_a $ on different asteroids are given in Fig. \ref{fig:acc_diagram-alltifactor1}. The fiducial model asteroid ($ D = 1 \km $) and the Phaethon-like model (Table \ref{tab: phys_phae}) are considered. These figures indicate that particles with radii from $ r_a < 1 $ to $ r_a \sim 10 \um $ tend to have larger accelerations outwards from AR. This holds for both our fiducial model asteroids ($ D = 1 \km $) and for Phaethon ($ D = 5.1 \km $).

As mentioned in Sect. \ref{res-temp prof}, the thermal inertia uncertainty does not induce a large uncertainty in $ \a_\mathrm{rad} \cdot \rhat $ during the day compared to the night. The difference is small between the perpendicular (solid) and aspect $ 45^\circ $ models (dashed), especially at night. The cause of the inward acceleration of the $ 1\, \mathrm{\mu m} $ olivine particle is given in Sect. \ref{disc-why 1um}.

\subsection{Conditions for outward acceleration} \label{res-size-size}
%%%%%%%%%%%%%%%%%%%%%%%%%%%%%%%%%%%%%%%%%%%%%%%%%%%%%%%%%%%%%%%%%%%%%%%%%%%%%%%%
In this section, the fictitious model asteroids are used to investigate the effects of some parameters. First, the effect of heliocentric distance $ \rh $ is vital to testing whether the outward dust acceleration is a phenomenon appearing universally on any asteroid or if it happens only in the near-Sun environment. Also, the effect of asteroid size $ D $ and rotational period $ P_\mathrm{rot} $ give hints as to whether the likelihood of losing small particles is dependent on the asteroid's size and rotation. Here, these parameters are investigated by simulating the particles on fictitious asteroids of varying $ D $, aspect angle, and $ P_\mathrm{rot} $, at several $ \rh $. Two particle types (magnetite and olivine) are investigated for varying radius $ r_a $. 

\begin{figure*}[t!]
  \centering
  \includegraphics[width=\linewidth]{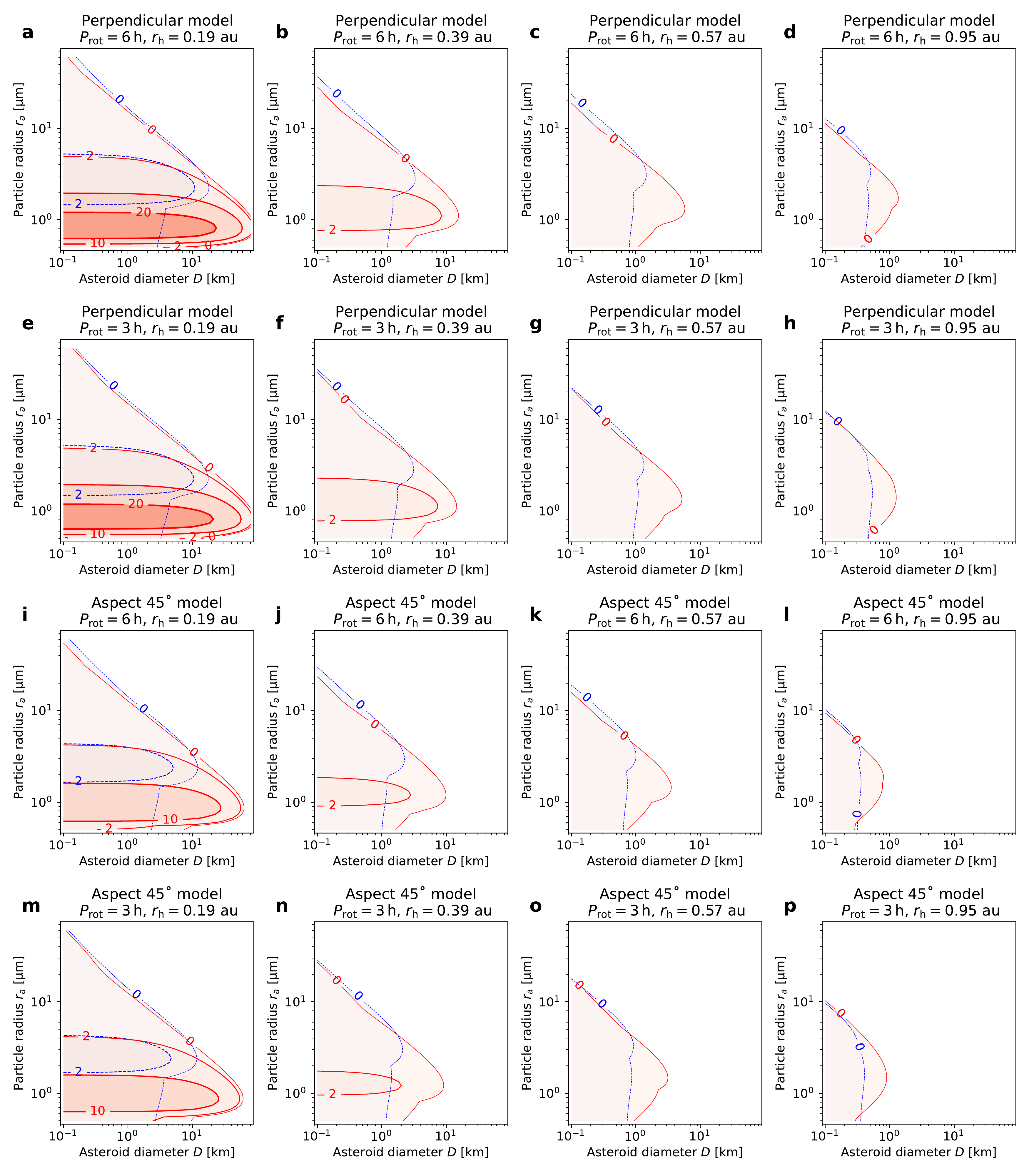}
  \caption{Maximum \textit{total}  radial acceleration, $ \mathbf{a}_\mathrm{total} \cdot \rhat = (\a_\mathrm{rad} + \a_\mathrm{grav} + \a_\mathrm{cen}) \cdot \rhat $ [$ \mathrm{mm/s^2} $] of particles on the equator of the fictitious model asteroids. Each row corresponds to a different asteroid model (\textbf{a} to \textbf{d}, perpendicular model with $ P_\mathrm{rot} = 6 \h $, \textbf{e} to \textbf{h}, perpendicular model with $ P_\mathrm{rot} = 3 \h $, \textbf{i} to \textbf{l}, aspect $ 45^\circ $ model with $ P_\mathrm{rot} = 6 \h $, and \textbf{m} to \textbf{p}, aspect $ 45^\circ $ model with $ P_\mathrm{rot} = 3 \h $). Each column corresponds to different heliocentric distances. The maximum acceleration is calculated by finding the maximum value along the longitude on the \textit{equator} for the given asteroid diameter $ D $ and particle radius $ r_a $. The \textcolor{blue}{blue} dashed lines and \textcolor{red}{red} solid lines indicate olivine and magnetite, respectively. The thickness of the contour increases as the acceleration value increases for fixed contour levels $ \mathbf{a}_\mathrm{total} \cdot \hat{\mathbf{r}} = \{0,\, 2,\, 10,\, 20,\, 40\} \,\mathrm{mm/s^2} $. Negative (inward) accelerations are not drawn.}
  \label{fig:size-size_all}
\end{figure*}

%At the initial state, i.e., a stationary particle at height $ H = 1\cm $ above the AR, the magnitude of the net radial radiative acceleration ($ \a_\mathrm{rad} \cdot \rhat $) depends on $ \rh $ but nearly independent of $ D $ (see Eqs. \ref{eq: acc_odot}, \ref{eq: acc_ref}, and \ref{eq: acc_ther}).
%; note $ \tilde{H} \approx 1 $ regardless of $ D $ at such small $ H $
%On the other hand, the threshold $ \| (\a_\mathrm{grav} + \a_\mathrm{cen}) \cdot \rhat \| = | a_\mathrm{cen} - a_\mathrm{grav} | $ is linearly proportional to $ D $. Therefore, 
The total radial acceleration at the initial state, that is, a stationary particle at height $ H = 1\cm $ above the AR, is expressed as a function of $ D $ and $ r_a $ at given $ \rh $ by $ \a_\mathrm{total} \cdot \rhat = \a_\mathrm{rad} \cdot \rhat + (\a_\mathrm{grav} + \a_\mathrm{cen}) \cdot \rhat $. The maximum of $ \a_\mathrm{total} \cdot \rhat $ along the equator is then found and plotted as a function of $ (D, r_a) $ for a few selected models (aspect angle, rotational period) and heliocentric distances, as in Fig. \ref{fig:size-size_all}.

From the figure, the maximum of $ \mathbf{a}_\mathrm{total} \cdot \rhat $ peaks at $ r_a \approx 1 \um $ for magnetite and at $ r_a \approx 2\mathrm{-}3 \um $ for olivine. The same trend must be true for realistic AR materials if they have $ \bar{Q}_\mathrm{pr} $ and $ \rho_m $ similar to these particles. There is a general trend that less particles can achieve outward net acceleration if $ \rh $ or $ D $ increases. The former is because $ T_S $ gets smaller as $ \rh $ increases (thus reducing $ a_\mathrm{ther} $ in Eq. (\ref{eq: acc_ther})), and the latter is because $ a_\mathrm{grav} $ (Eq. (\ref{eq: a_grav})) dominates the radial total acceleration. 

No large change is visible for models with different $ P_\mathrm{rot} $ in Fig. \ref{fig:size-size_all}. As mentioned in Sect. \ref{meth-tpm}, the temperature on AR after halving $ P_\mathrm{rot} $ is identical to that when $ \Gamma = 200 \tiu $ is updated to $ \sqrt{2} \Gamma = 283 \tiu $, because both $ \Theta $ and $ T_\mathrm{eqm} $ are the same for these two cases. However, such a small change in $ \Gamma $ does not result in a large change in the temperature profile at noon in the near-Sun environment (Fig. \ref{fig:tempprof}). As the maximum $ \a_\mathrm{total} \cdot \rhat $ appear near noon (Fig. \ref{fig:acc_diagram-alltifactor1}), it is expected that the change in $ P_\mathrm{rot} $ by a factor of two will not result in a large change in $ \a_\mathrm{rad} $ and thus in $ \a_\mathrm{total} $. Therefore, the change in $ P_\mathrm{rot} $ does not change the results shown in Fig. \ref{fig:size-size_all} significantly. 

Comparing the rows in Fig. \ref{fig:size-size_all}, it is often seen that particles on $ P_\mathrm{rot} = 6\h $ have slightly greater outward acceleration than $ P_\mathrm{rot} = 3\h $. This result may contradict an intuitive prediction that centrifugal acceleration of a fast rotator is larger than that of a slow rotator. A change in $ \Theta $ when changing $ P_\mathrm{rot} $ (as mentioned above, $ P_\mathrm{rot} \rightarrow 0.5 P_\mathrm{rot} $ has identical effect as $ \Gamma \rightarrow \sqrt{2}\Gamma $) slightly changes the maximum temperature as in Fig. \ref{fig:tempprof}. Hence, $ a_\mathrm{rad} $, which is a sensitive function of the temperature (Eq. (\ref{eq: acc_ther})) is reduced near noon, where the maximum outward acceleration likely happens. As these two effects take place at the same time, the contours in Fig. \ref{fig:size-size_all} are not necessarily expected to show a simple trend as a function of $ P_\mathrm{rot} $.

Also, in Fig. \ref{fig:size-size_all}, some bumps are seen especially at $ \mathrm{max} \{ \a_\mathrm{total} \cdot \rhat \} = 0 $ contours around $ r_a \sim 1 \um $. Figure \ref{fig:acc_diagram-alltifactor1}\textbf{a} is an example for understanding the olivine (blue dashed) contour in Fig. \ref{fig:size-size_all}\textbf{a}. In the figure, the location where $ \mathrm{max} \{ \a_\mathrm{total} \cdot \rhat \} $ occurs changes from the evening terminator (when $ r_a \lesssim 1\um $) to noon (when $ r_a \gtrsim 1.5 \um $). Therefore, the olivine contour in Fig. \ref{fig:size-size_all}\textbf{a} will change its trend at around $ r_a \sim 1\um $.

\begin{figure}
\centering
\includegraphics[width=\linewidth]{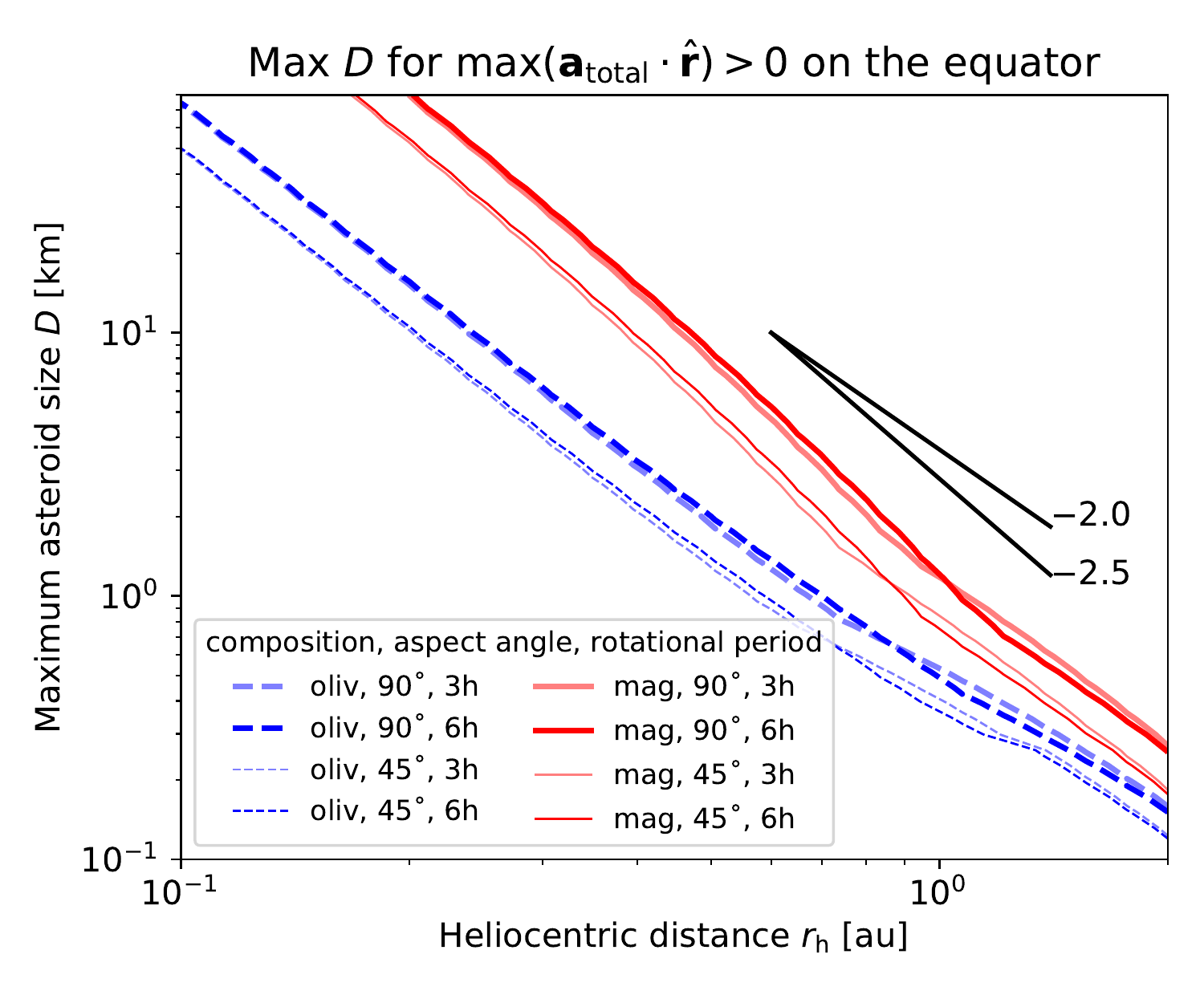}
\caption{Maximum diameters of asteroids ($ D $) that produce positive $ \mathrm{max}(\a_\mathrm{total} \cdot \rhat) $ on the equator as a function of heliocentric distance ($ \rh \in [0.1, 2.0] \au $). The \textcolor{red}{red} solid and \textcolor{blue}{blue} dashed lines are the results for magnetite and olivine, respectively. The results of perpendicular and aspect $ 45^\circ $ models are discriminated by the line thicknesses, and the results of different rotational periods ($ 6\h $ and $ 3\h $) are discriminated by the transparency of the lines. The black lines show the slope according to a power law.}
\label{fig:rhdiammax}
\end{figure}

Finally, using the calculated $ \mathrm{max}(\a_\mathrm{total} \cdot \rhat) $ values, we constrain the diameter of asteroids that can accelerate particles. As seen in Fig. \ref{fig:size-size_all}, the condition for diameter gives only the upper bound under the parameter space considered in this study, and hence the maximum diameter is plotted in Fig. \ref{fig:rhdiammax}. The derived maximum diameters of asteroids are smaller for olivine than magnetite, because olivine is less affected by the thermal radiation (smaller $ \bar{Q}_\mathrm{pr}(r_a, T_S) / \bar{Q}_\mathrm{pr, \odot}(r_a) $ in Fig. \ref{fig:qprbar}). Also, the clear trend of decreasing maximum diameter is seen as $ \rh $ increases, that is, as the thermal radiation weakens. The profiles in Fig. \ref{fig:rhdiammax} approximately follow a power law $ \rh^{-(2\mathrm{-}2.5) } $, as indicated by the black lines in the figure. At $ \rh = 1 \au $, particles can be accelerated outward to space only if $ D \lesssim 1\km $ under the model used in this study.

Summarizing the above results in this section, we find the following key points: (1) the thermal radiation becomes critical as an asteroid approaches the Sun, (2) smaller asteroids tend to lose particles more easily than larger asteroids, (3) particles of $ r_a \sim 1\um $ are most efficiently accelerated outwards, and (4) these results are less sensitive to rotational period unless $ a_\mathrm{cen} \gtrsim a_\mathrm{grav} $.

\subsection{Phaethon}\label{res-phaethon}
%%%%%%%%%%%%%%%%%%%%%%%%%%%%%%%%%%%%%%%%%%%%%%%%%%%%%%%%%%%%%%%%%%%%%%%%%%%%%%%%
Asteroid (3200) Phaethon is the most appropriate object with which to test the validity of our model because it has shown activity near the perihelion \citep{2010AJ....140.1519J, 2013AJ....145..154L, 2013ApJ...771L..36J, 2017AJ....153...23H} but has not shown lingering activity at $ \rh \sim 1 \au $ \citep{1985MNRAS.214P..29G, 1992Icar...97..276L, 1996Icar..119..173C, 2005ApJ...624.1093H, 2008Icar..194..843W, 2018AJ....156..238J, 2019AJ....157..193J}. 
% 1984Icar...59..296C : gas emission not nebulosity
If the recurrent near-perihelion activities are caused by the same mechanism over time, any model explaining the phenomenon must coincide with the observations, in that: (A) Phaethon's activity (dust ejection) must be stronger near the perihelion, and (B) the ejected particles must have a spherical equivalent radius of $ r_a \sim 1 \um $ \citep{2013ApJ...771L..36J, 2017AJ....153...23H}. 

\begin{figure}
  \centering
  \includegraphics[width=\linewidth]{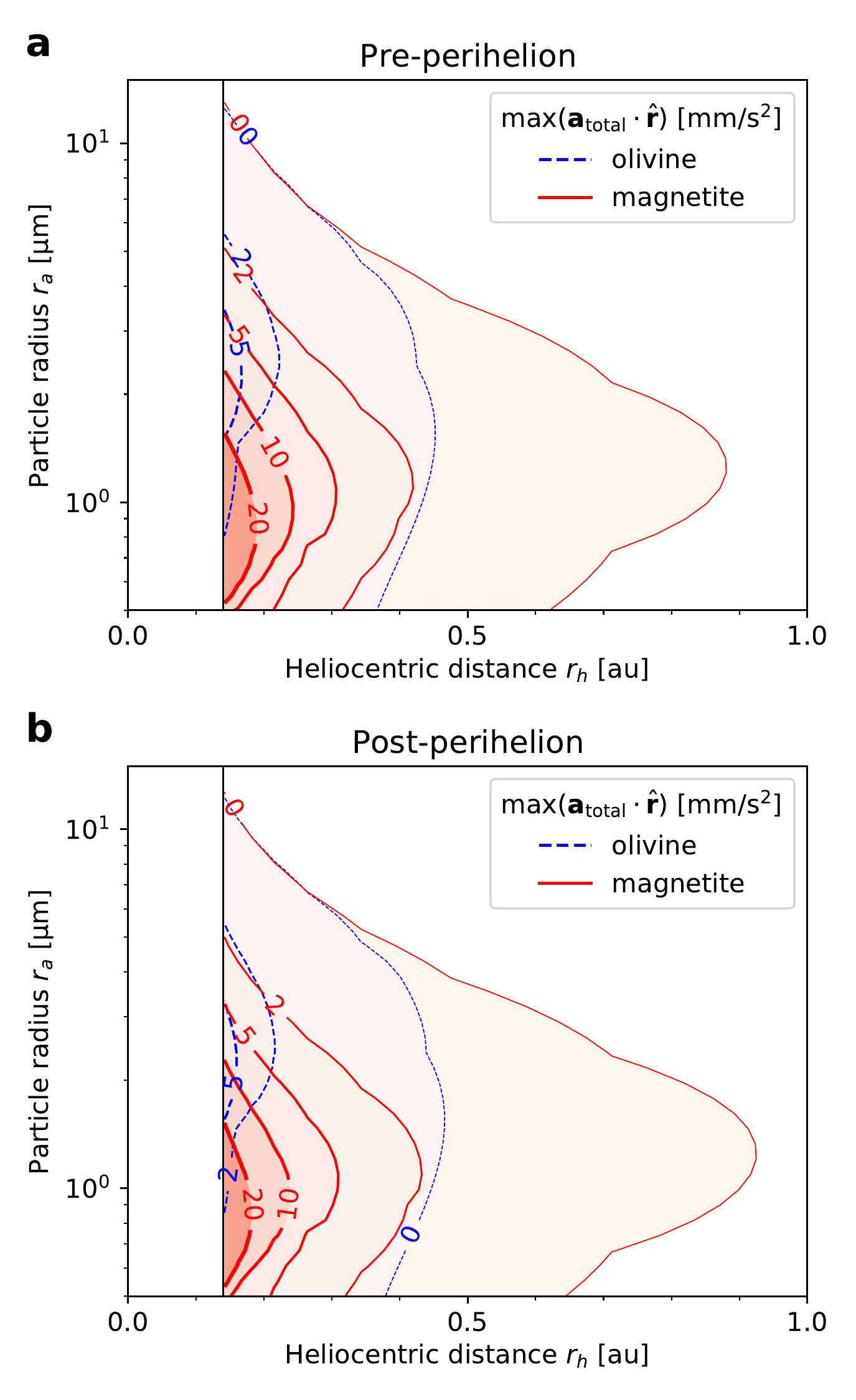}
  \caption{Maximum radial acceleration on the Phaethon's \textit{equator}. As in Fig. \ref{fig:size-size_all}, the maximum acceleration on the equator is calculated for a wide range of particle sizes ($ 0.5 \mathrm{-} 40 \um$) before (\textbf{a}) and after (\textbf{b}) the perihelion passage. The black vertical line corresponds to the Phaethon's perihelion. The numbers on the contour plots are in units of $ \mathrm{mm/s^2} $, and the thickness of the contour increases as the acceleration value increases for fixed contour levels $ \mathbf{a}_\mathrm{total} \cdot \hat{\mathbf{r}} = \{0,\, 2,\, 5,\, 10,\, 20\} \,\mathrm{mm/s^2} $. The \textcolor{blue}{blue} dashed and \textcolor{red}{red} solid lines indicate olivine and magnetite, respectively.}
  \label{fig:phaeacceq}
\end{figure}

To test the ability of our model to fulfil the requirements, the maximum $ \mathbf{a}_\mathrm{total} \cdot \rhat $ was found, as in Sect. \ref{res-size-size}, along Phaethon's equator. The physical parameters are as given in Table \ref{tab: phys_phae}. The results are shown in Fig. \ref{fig:phaeacceq} as a function of heliocentric distance $ \rh $ and particle radius $ r_a $ for magnetite and olivine. If the maximum $ \mathbf{a}_\mathrm{total} \cdot \rhat $ on the equator can be a proxy of the likelihood of dust ejection, this figure tells us that Phaethon should show the strongest dust ejection near the perihelion. Over time, as Phaethon's $ \rh $ increases (Sect. \ref{disc-phae_orbit} and Fig. \ref{fig:phaeorbit}), the likelihood will rapidly drop, and eventually converge to zero when $ \rh \gtrsim 0.8 \au $. Therefore, our model satisfies the requirement (A).

In addition, Fig. \ref{fig:phaeacceq} indicates that particles of radius from $ r_a < 1 \um $ to $ r_a \sim 2\mathrm{-}3 \um $ are likely to have larger accelerations outwards from the AR, depending on the material. This is similar to the results shown in Sects. \ref{res-acc prof} (Fig. \ref{fig:acc_diagram-alltifactor1}) and \ref{res-size-size} (Fig. \ref{fig:size-size_all}), which used fictitious model asteroids. Moreover, this is highly consistent with observational studies, as in requirement (B). Therefore, we conclude that our model satisfies the two important requirements outlined above and based on previous observational studies of Phaethon.

The two plots in Fig. \ref{fig:phaeacceq} (pre- and post-perihelion) appear similar because, by coincidence, the aspect angle of Phaethon at perihelion is almost $ 90^\circ $ (strictly speaking, $ 101^\circ $). This means the geometric configuration of the Sun and AR patch on the equator are nearly time-symmetric with respect to the perihelion.

%%%%%%%%%%%%%%%%%%%%%%%%%%%%%%%%%%%%%%%%%%%%%%%%%%%%%%%%%%%%%%%%%%%%%%%%%%%%%%%%
\section{Discussion}
%%%%%%%%%%%%%%%%%%%%%%%%%%%%%%%%%%%%%%%%%%%%%%%%%%%%%%%%%%%%%%%%%%%%%%%%%%%%%%%%

\subsection{The acceleration peak at $ r_a \sim 1  \um $} \label{disc-why 1um}
%%%%%%%%%%%%%%%%%%%%%%%%%%%%%%%%%%%%%%%%%%%%%%%%%%%%%%%%%%%%%%%%%%%%%%%%%%%%%%%%

In this section, we provide qualitative discussions regarding the result that $ r_a \sim 1 \um $ particles have the largest acceleration.
Larger particles tend to have negative $ \a_\mathrm{total} \cdot \rhat $ values because gravity dominates $ \a_\mathrm{total} $, as the area-to-mass-ratio decreases. The coefficient $ \bar{Q}_\mathrm{pr} $ plays a key role, especially for small particles. In Fig. \ref{fig:qprbar}, it is seen that $ R_Q = \bar{Q}_\mathrm{pr} (T_S, r_a) /\bar{Q}_\mathrm{pr, \odot} (r_a) $ gets small for smaller particle sizes. This means that solar radiation dominates the thermal radiation, and hence the particle is easily pushed into the AR.

%Besides, both the upper and lower limits are set by the $ \bar{Q}_\mathrm{pr} $ (Eq. (\ref{eq: bar Qpr def})), the \textit{efficiency} of the particle being accelerated by radiation (Sect. \ref{meth-mie}). It is a nonlinear function of $ r_a $ and the blackbody temperature of the incident spectrum, as shown in Fig. \ref{fig:qprbar}. Therefore, the dominancy between the solar and thermal radiations, $ a_\mathrm{ther}/a_\odot $, is also a nonlinear function of $ r_a $ as in Eq. (\ref{eq: a_ther/a_odot z}). For a particle on AR, the ratio in Eq. (\ref{eq: a_ther/a_odot z}) is mainly controlled by the relative efficacy $ R_Q (T_S, r_a) := \bar{Q}_\mathrm{pr} (T_S, r_a) /\bar{Q}_\mathrm{pr} (T_\odot, r_a) $ (Fig \ref{fig:RQplot}). $ R_Q $ has a peak at approximately $ r_a = \mathrm{few} \um $ for $ T = 1000 \,\mathrm{K} $ regardless of the composition (olivine and magnetite). The difference in the $ R_Q $ comes from the different optical indices of the materials.

This is why particles of $ r_a \sim 1 \um $ tend to experience outward acceleration, while larger or smaller particles tend to experience weaker, or even inward acceleration ($ \mathbf{a}_\mathrm{total} \cdot \rhat < 0 $). For olivine, the ratio $ R_Q $ decreases steeply so that solar radiation dominates $ \mathbf{a}_\mathrm{total} \cdot \rhat $ during the daytime. For this reason, the olivine particle of $ r_a = 1 \um $ in Fig. \ref{fig:acc_diagram-alltifactor1} displays an inward acceleration during the daytime.

\subsection{Effect of thermal inertia $ \Gamma $} \label{disc-effect of TI}
%%%%%%%%%%%%%%%%%%%%%%%%%%%%%%%%%%%%%%%%%%%%%%%%%%%%%%%%%%%%%%%%%%%%%%%%%%%%%%%%

\cite{2007Icar..190..236D} and \cite{2015aste.book..107D} discuss the effect of the temperature dependency of $ \Gamma $. The dependency is caused mostly by the temperature dependency of the thermal conductivity, $ \kappa $. Although $ \Gamma $ must change at each voxel in TPM (depending on the temperature at each position on the asteroid and depth), a first-order approximation to cope with this temperature dependence is to roughly scale $ \Gamma $ by $ \Gamma \propto T_S^{3/2} \propto \rh^{-3/4} $ or 
\begin{equation}\label{eq: TI rh}
  \Gamma(\rh) = \left ( \frac{\rh}{1 \au} \right ) ^{-3/4} \Gamma_1 
,\end{equation}
where $ \Gamma_1 $ is the thermal inertia at $ 1 \au $ \citep{2015aste.book..107D}. In the present work, the thermal inertia of the  model asteroids is taken to be $ 200\tiu $ from the $ \Gamma_1 $ values of the similar-sized asteroids \cite[][Figure 9]{2015aste.book..107D}. If the asteroids were placed at $ \rh = 0.2 \au $, the true effective thermal inertia would be $ \Gamma = 669 \tiu $ following this relationship. Phaethon's thermal inertia is also determined at $ \rh \sim 1\au $ \citep{2016A&A...592A..34H, 2018A&A...620L...8H}, so the value at its perihelion, $ \rh = 0.14 \au $, is $ \Gamma = 2622\tiu$. The temperature profiles using these $ \Gamma $ values are plotted in Fig. \ref{fig:tempprof}.

\begin{figure*}[t!]
  \centering
  \includegraphics[width=\linewidth]{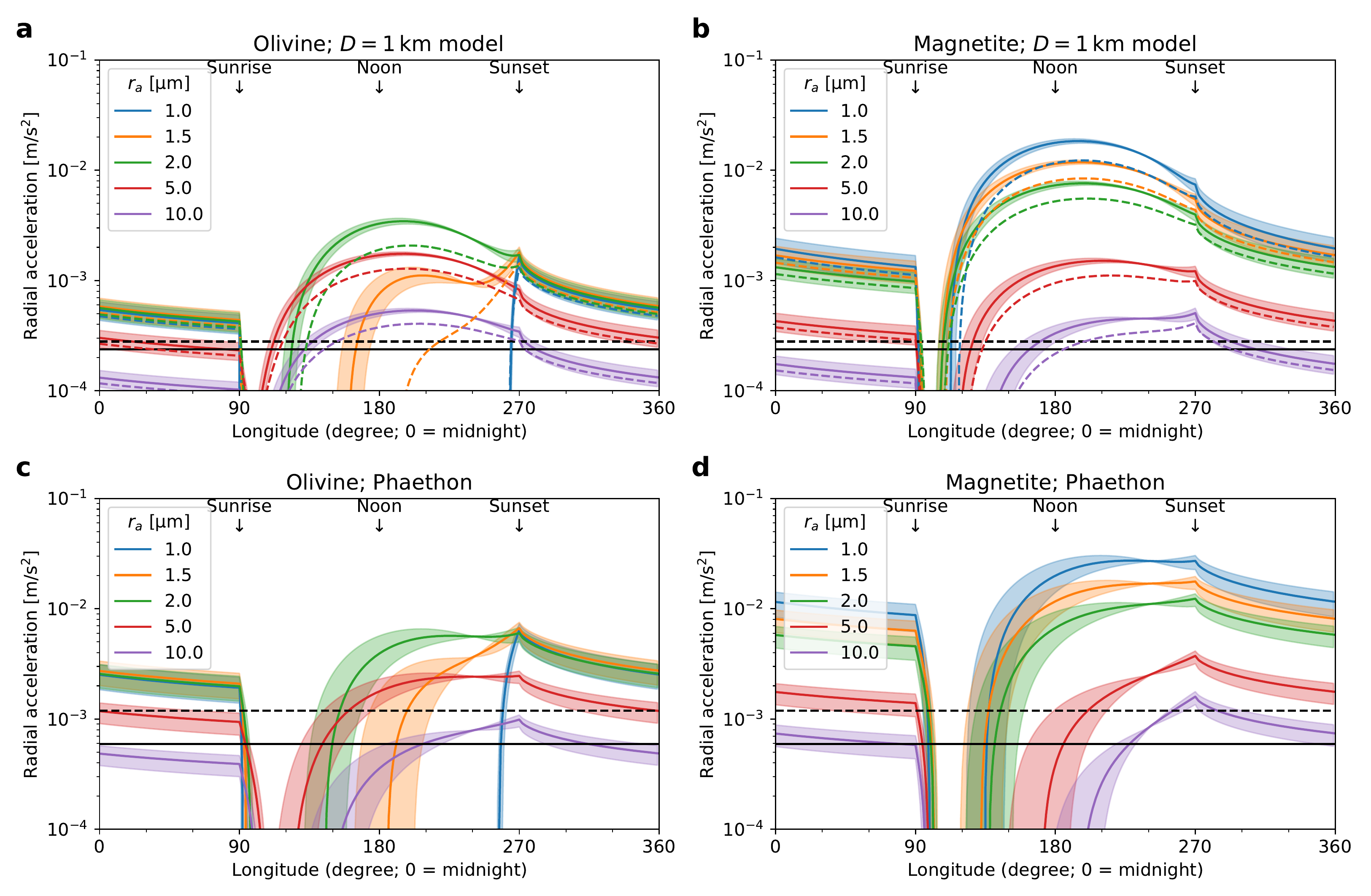}
  \caption{Identical plot to Fig. \ref{fig:acc_diagram-alltifactor1}, except the thermal inertia of asteroids is changed according to Eq. (\ref{eq: TI rh}). \textbf{a, c}, olivine; \textbf{b, d}, magnetite. The shade in each plot signifies the uncertainty on thermal inertia, scaled by the same factor, $ (\rh / 1 \au)^{-3/4} $.}
  \label{fig:accdiagram-alltifactor5}
\end{figure*}

Updating the thermal inertia values of the model asteroids and Phaethon to these larger values, we obtain Fig. \ref{fig:accdiagram-alltifactor5}. When the thermal inertia $ \Gamma $ is updated from the fiducial value, the morning and noon temperatures decrease because the heating process takes longer. On the other hand, the afternoon and night temperatures increase because cooling also takes longer, as depicted in Fig. \ref{fig:tempprof}. This implies that the $ a_\mathrm{ther} $ for the models with increased $ \Gamma $ will be larger than fiducial models in the afternoon. This effect is confirmed by comparing Figs. \ref{fig:acc_diagram-alltifactor1} and \ref{fig:accdiagram-alltifactor5}.

%In Fig. \ref{fig:tempprof}, the daytime temperature profile does not change significantly over a wide range of thermal inertia, compared to the cases of $ \rh \sim 1\au $ (e.g., Fig 2.2 of \citealt{MuellerM2007PhDT}). The reason can be inferred from Eq. (\ref{eq: theta-def}): 
%\begin{equation}
%  \Theta \propto \Gamma T_\mathrm{eqm}^{-3} \propto \Gamma \rh^{3/2} \propto \Gamma_1 \rh^{3/4} ~,
%\end{equation}
%where Eq. (\ref{eq: TI rh}) is used. This indicates that the change in absolute value of $ \Theta $ is smaller at smaller $ \rh $ for a given change in $ \Gamma $ or $ \Gamma_1 $. As $ \Theta $ is the only parameter which tunes the shape of temperature profiles (Sect. \ref{meth-tpm}), this means the temperature profile changes milder at small $ \rh $ under the change in $ \Gamma $. It can also be understood from the surface energy balance (Sect. \ref{meth-tpm}; energy radiated from AR surface $ = $ energy conducted from subsurface to surface $ + $ incident solar energy): The thermal conduction term gets negligible compared to the incident solar energy term at small $ \rh $. Therefore, the daytime temperature profile gets less sensitive to $ \Gamma $ in the near-Sun environment. In the evening or at night, however, $ \Gamma $ changes the temperature profile significantly. 

\subsection{Possible mechanisms to generate and lift particles to initial height}\label{initial condition}
%%%%%%%%%%%%%%%%%%%%%%%%%%%%%%%%%%%%%%%%%%%%%%%%%%%%%%%%%%%%%%%%%%%%%%%%%%%%%%%%
%So far, we did not discuss how the initial conditions are achieved, i.e., how particles could be generated and afloat above the AR at first. Although these are not the main focus of this work, we briefly discuss few possible mechanisms to justify our assumptions on initial condition settings. 

The accelerations shown in this work, such as the radiative accelerations, have a maximum order of $ < 100 \,\mathrm{mm/s^2} $. However, this acceleration is still too small to overcome the cohesion force of the regolith particles. According to Eq (12) of \cite{2011P&SS...59.1758H}, the acceleration equivalent to the cohesion force is  

\begin{equation}\label{eq: a_co}
  a_{co} = 4.09\times 10^8 
    \frac{\left ( \frac{S}{0.1} \right )^2}
    {\left (\frac{\rho_m}{3000 \den} \right )
    \left (\frac{r_a}{1 \um} \right )^{2}} \,\mathrm{[mm/s^2]} ~,
\end{equation}
where $ S = 0.1\mathrm{-}1 $ is the `cleaness' parameter and the cohesion force in the original work is divided by particle mass $ m = 4\pi \rho_m r_a^3 / 3 $. Thus, an event to break this cohesion force is likely to have happened to free the particle from the regolith and achieve the required initial condition $ H \gg r_a $. Two such mechanisms are briefly discussed, and then the implication to Phaethon's orbit is discussed.

\subsubsection{Thermal fatigue} \label{disc-thermal fatigue}
%-------------------------------------------------------------------------------
\cite{2019JGRE..124.3304R} provided a simple visualization of two (insufficient, as they mention, yet useful) criteria to investigate the probability of thermal fatigue (Figure 5 of their work). First is $ \Delta \hat{T}_S > 100\,\mathrm{K,} $ where $ \Delta \hat{T}_S $ is the daily maximum temperature difference on the asteroid's surface, that is, the amplitude of the curves in Fig. \ref{fig:tempprof}. The second criterion is $ \left | d\hat{T}_s / dt \right | > 2 \,\mathrm{K/min}$, where $ \left | d\hat{T}_S / dt \right | $ is the maximum time-rate of change in surface temperatures, that is, the maximum time derivative of the temperature profile. In the original work, the thermal inertia and Bond albedo of the asteroids were fixed while rotational period and heliocentric distance are changed. 

\begin{figure*}[t!]
  \centering
  \sidecaption
  \includegraphics[width=0.75\linewidth]{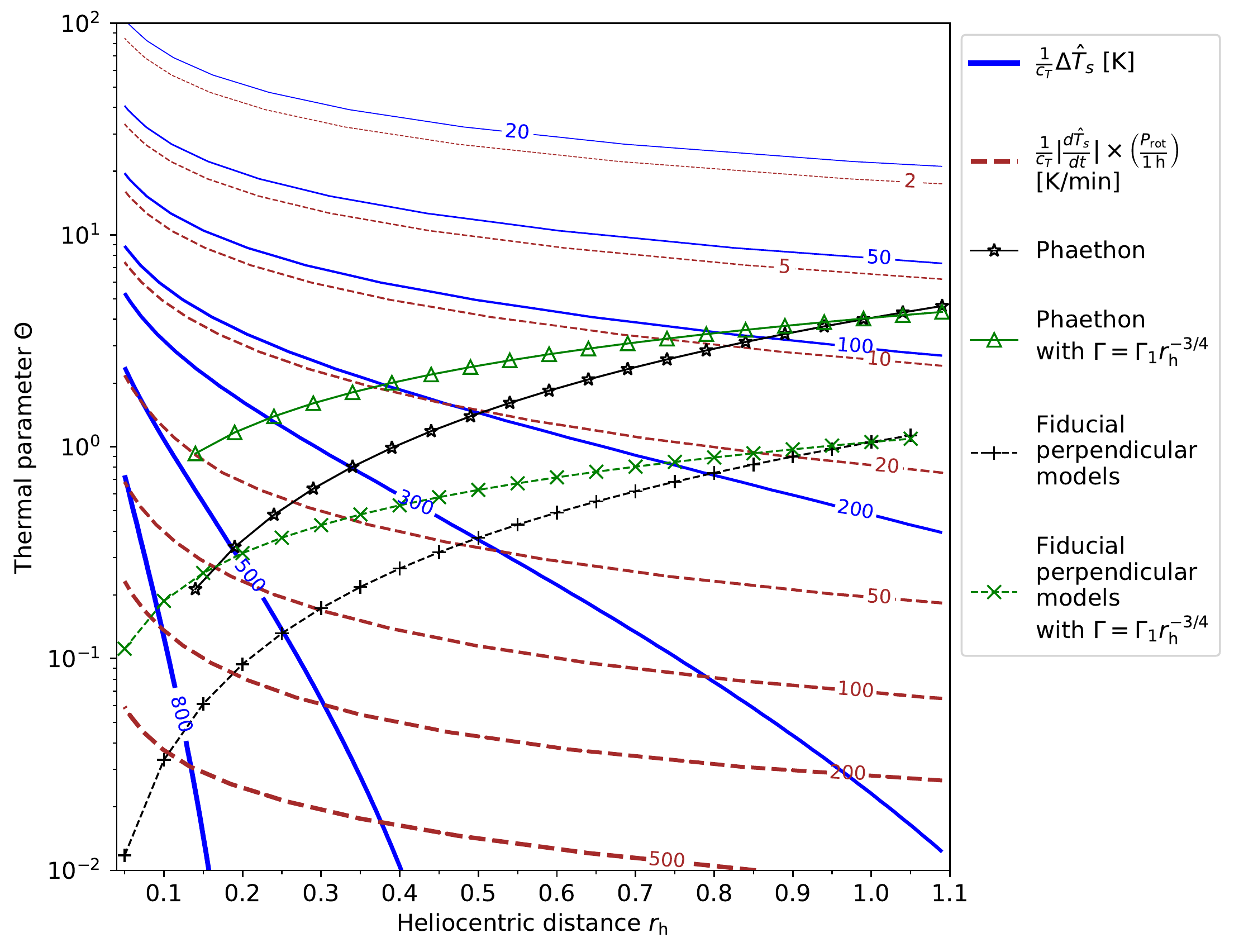}
  \caption{Thermal fatigue plot. The thermal parameter ($ \Theta $) over heliocentric distance ($ \rh $) for Phaethon (solid lines, $ \rh \ge 0.14  \au$) and fiducial perpendicular model asteroids with varying $ \rh $ (dashed lines) are shown. The black star and plus symbols indicate that $ \Gamma $ is fixed, while the green triangle and cross markers show results where $ \Gamma $ is assumed to be changed by Eq. (\ref{eq: TI rh}). The blue solid lines are the contours of $ \frac{1}{c_T} \Delta \hat{T}_S $, and the brown dashed lines are that of $ \frac{1}{c_T} \left | \frac{d\hat{T}_S}{dt} \right | \times \left ( \frac{P_\mathrm{rot}}{1 \h} \right ) $. The values shown for the contours are the $ \Delta \hat{T}_S $ and $ \left | \frac{d\hat{T}_S}{dt} \right | $ values for the case when $ c_T = 1 $ and $ P_\mathrm{rot} = 1 \h$. The contours get thicker for larger values to guide the eyes.}
  \label{fig:theta-rh}
\end{figure*}

We devise a similar but more general plot as shown in Fig. \ref{fig:theta-rh}. This figure visualizes the likelihood of thermal fatigue on general  small bodies of the Solar System. Taking the Bond albedo, the emissivity, and the rotational period as free parameters, the two criteria are re-formulated by  
\begin{equation} \label{eq: thermal_fatigue_1}
  \frac{1}{c_T} \Delta \hat{T}_S > \frac{100 \,\mathrm{K}}{c_T}
,\end{equation}
and 
\begin{equation}\label{eq: thermal_fatigue_2}
  \frac{(P_\mathrm{rot} / 1\h)}{c_T} \left | \frac{d\hat{T}_S}{dt} \right | 
    > \frac{(P_\mathrm{rot} / 1\h)}{c_T} \times 2\,\mathrm{K/min} ~. 
\end{equation}
For a detailed derivation, see Appendix \ref{app: fatigue reparam}.

The fiducial perpendicular model and Phaethon have $ (c_T, P_\mathrm{rot}) = (1.014, 6\h) $ and $ (1.015, 3.604\h) $, and so the right-hand side of Eq. (\ref{eq: thermal_fatigue_1}) is $ 98.7 \,\mathrm{[K]} $ and $ 98.6 \,\mathrm{[K]} $, respectively. This criterion can be checked by tracing blue solid contours in Fig. \ref{fig:theta-rh}. The right-hand side of Eq. (\ref{eq: thermal_fatigue_2}) becomes $ 11.8 \,\mathrm{[K/min]}$ and $ 7.1 \,\mathrm{[K/min]} $, respectively (brown dashed contours in Fig. \ref{fig:theta-rh}). From the figure, the fiducial model at $ \rh \lesssim 1.5 \au $ and Phaethon at $ r_h \lesssim 0.8 \au $ satisfy both of the criteria, and hence, thermal fatigue is likely to occur in those regions (Region III in the notation of \citealt{2019JGRE..124.3304R}). The \textit{probability} of the occurrence of the fracture and the initial size and velocity distribution of the generated particle are beyond the scope of this work. Previous work \citep{2010AJ....140.1519J} suggested that thermal expansion potential energy is translated into kinetic energy and can accelerate the particle to a certain initial velocity immediately after the thermal fatigue.

\subsubsection{Electrostatic dust lofting and levitation} \label{disc-electrolofting}
%-------------------------------------------------------------------------------

The electrostatic lofting mechanism, which is the mechanism that launches the particles from the AR, has recently been studied extensively both theoretically \citep{2016JGRE..121.2150Z} and experimentally \citep{2016GeoRL..43.6103W,2018GeoRL..4513206H}. If small particles are residing on the rocks by cohesion force, a supercharging effect near the terminator or the shadow on asteroids may free the particle from the AR with initial kinetic energy and achieve the condition $ H \gg r_a $. 

From \cite{2016GeoRL..43.6103W} (Figure 4 of them) and \cite{2018GeoRL..4513206H}, particles of radius $ r_a = 5 \mathrm{-} 22 \um $ ($ 2 r_a = 10 \mathrm{-} 44 \um $) could achieve an initial speed of up to $ v_{z, 0} \sim 0.6 \,\mathrm{m/s} $ due to the supercharging effect at $ \rh = 1 \au $. This is the aftermath of Coulomb interaction, and therefore the particle size range and initial speed are independent of asteroid size ($ D $) or $ \rh $. %$ \rh $ only changes the rate of lofting, not the initial speed or loftable particle size ranges. 
Thus, electrostatic dust lofting due to the supercharging effect can achieve the initial condition ($ H \gg r_a $).

It is found that the dust lofting mechanism does not repeat forever on the asteroid; rather, the dust lofting rate decreases over time during the experiment \citep{2018GeoRL..4513206H}. Therefore, it is expected that the electrostatic dust lofting will occur only immediately after the AR is \textit{reset}, that is, when fresh small particles are generated by micrometeoritic bombardment or thermal fatigue. 

\textsl{}

\subsubsection{Phaethon's orbital movement} \label{disc-phae_orbit}
\begin{figure}
  \centering
  \includegraphics[width=\linewidth]{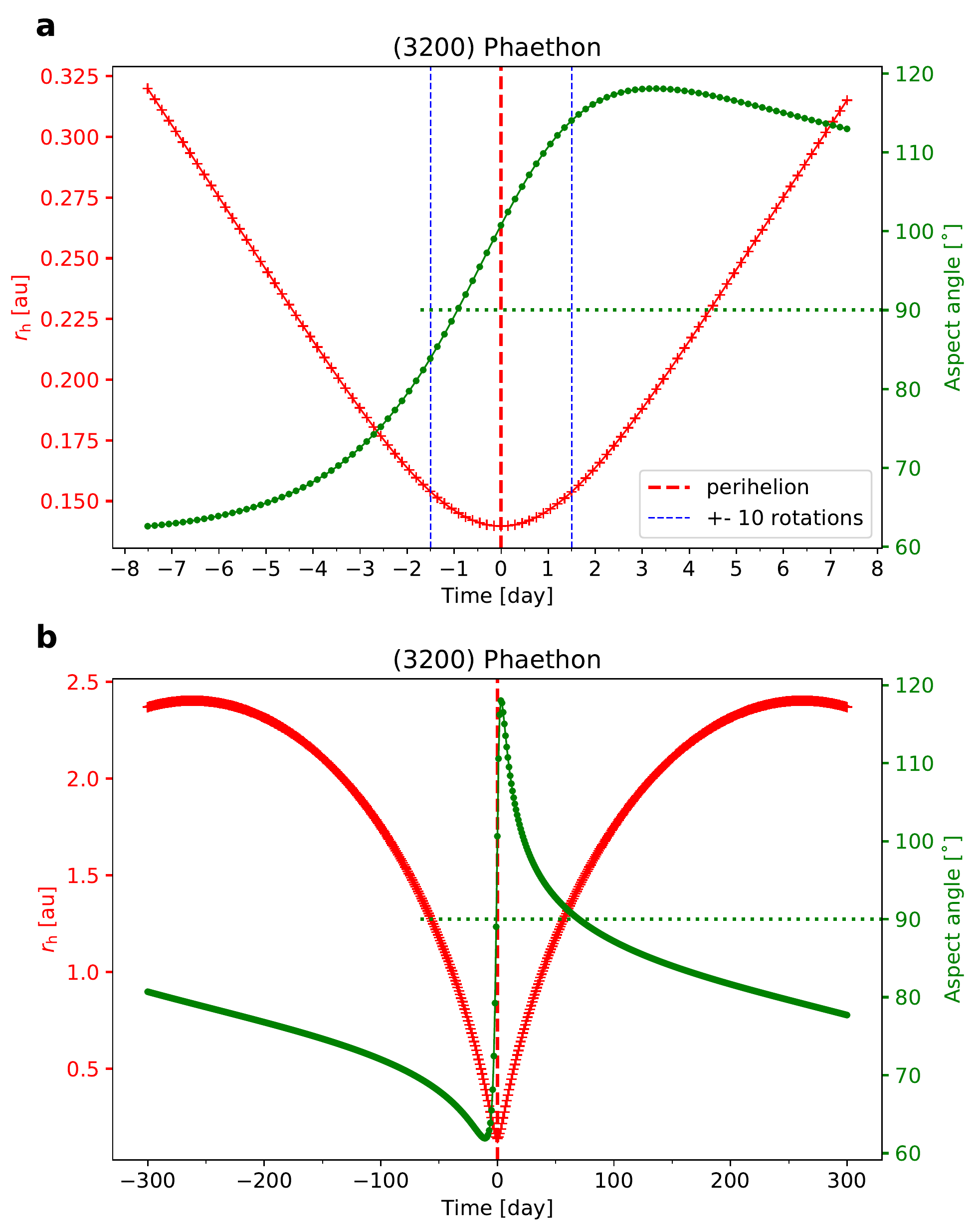}
  \caption{Heliocentric distance (red plus; left ordinate) and the aspect angle (green dot; right ordinate) of asteroid (3200) Phaethon over time. \textbf{a}: 50 rotations before and after the perihelion; \textbf{b}: over one full orbit. Negative and positive times denote pre- and post-perihelion, respectively, where perihelion is indicated by the thick red dashed vertical line. The thin blue dashed vertical line in \textbf{a} indicates $ \pm 10 P_\mathrm{rot} $ of Phaethon from its perihelion. The markers in \textbf{a} are separated by $ P_\mathrm{rot} $ of Phaethon, while those in \textbf{b} are separated by 1 day. We note that the heliocentric distance changes by $ \sim 10\,\% $ over ten rotations, and the aspect angle changes by $ \sim 1.5^\circ $ per rotation near the perihelion.}
  \label{fig:phaeorbit}
\end{figure}

%-------------------------------------------------------------------------------
As described in Sect. \ref{meth-tpm}, the TPM calculates the equilibrium temperature after rotating the asteroid at least 50 times and ignoring seasonal variation. Phaethon has a large eccentricity of $ 0.89 $, and therefore the true anomaly changes dramatically near its perihelion. Figure \ref{fig:phaeorbit} shows the change in $ \rh $ of Phaethon and the aspect angle using the pole orientation (Table \ref{tab: phys_phae}). During the ten asteroidal days ($ \approx 1.5 \,\mathrm{day} $) near the perihelion, $ \rh $ changes by $ \sim 10\,\% $ and aspect angle changes $ > 10^\circ $. The environment of Phaethon at its perihelion is a chaos of thermal \textit{in}equilibria under strong solar irradiation.

Although this weakens our assumption that the seasonal effect is negligible, such information can provide important hints as to the probability of thermal fatigue and electrostatic dust lofting. The chaotic inequilibria result in a large and sudden change in temperature not only along the longitude but also along the latitude due to the quick change in the aspect angle. As soon as fracture happens and small particles are generated, they may have reached an initial height or be lofted by the supercharging effect. In this work, initial velocity was set to zero with a small initial height above AR ($ H = 1 \cm $) to show the validity of our idea even when the initial condition is unfavorable. If the particles have a positive initial speed outwards, more particles could have ejected. Moreover, there is a time-symmetry in the illumination condition on Phaethon along the latitude: The thermal (illumination) history of the north during the perihelion passage is nearly the time-reversal of that of the south because the aspect angle is nearly $ 90^\circ $. This may have implications on the latitudinal particle size distribution on Phaethon, but investigation of this matter is beyond the scope of this work.

% Detailed description (removed)
%As Phaethon approaches its perihelion, for example, aspect angle quickly increases from $ \sim 60^\circ $ to $ 100^\circ $, and southern part (latitude $ 90\mathrm{-} 60^\circ \mathrm{S}$) is suddenly illuminated. As Phaethon leaves perihelion, aspect angle still increases up to $ \sim 120^\circ $, leaving northern part (latitude $ 60\mathrm{-}90^\circ \mathrm{N} $) in the shadow, which were severely heated by the Sun only about a week ($ \sim 50 $ rotations) ago. 

\subsection{Particle trajectory integration} \label{disc-trajectory}
%%%%%%%%%%%%%%%%%%%%%%%%%%%%%%%%%%%%%%%%%%%%%%%%%%%%%%%%%%%%%%%%%%%%%%%%%%%%%%%%
One of the most important points to improve in this work is that the large acceleration does not necessarily guarantee the escape of particles. This problem is at least two-fold. First, the particle trajectory integration must be performed for varying height of the particle, until at least the Hill radius ($r_H = 66\,\mathrm{km}$; \citealt{2019AJ....157..193J}). Second, even though the particle can be ejected by thermal radiation, it must have been generated (by, e.g., micrometeorite bombardment or thermal fatigue), and the initial conditions must be known accurately. This second part is beyond the scope of this work and has already been discussed in previous sections to some extent, but we develop the first point below. 

We consider the particle trajectories in a frame in which the observer orbits around the Sun with Phaethon, but does not rotate around the asteroid. Thus, the particles are set to have an initial speed of $ v_0 = \pi D / P_\mathrm{rot} $ along the direction of the asteroid's rotation. Also, the centrifugal acceleration from the orbital motion and the gravitational acceleration of the Sun are balanced. In the given reference frame, Coriolis force is negligible.
% The Coriolis acceleration is proportional to the orbital angular speed, and the angular speed is specific angular momentum divided by $ \rh^2 $: 
%\begin{equation}
%  \omega_\mathrm{orb} = \frac{\sqrt{GM_\odot \mathcal{A} (1-e^2)}}{\rh^2} ~,
%\end{equation}
%where $ \mathcal{A} $ and $ e $ are the semi-major axis and eccentricity of the asteroid orbit, respectively. For Phaethon, $ \mathcal{A} = 1.27\au$ and $ e = 0.89 $ result in $ \omega_\mathrm{orb} = 5.2 \times 10^{-6} \,\mathrm{s^{-1}} $ at the perihelion. Therefore, the Coriolis acceleration is only about the order of \textit{maximum} order of $ a_\mathrm{Coriolis, max} \sim 2 \omega_\mathrm{orb} v_\mathrm{esc} \approx 10^{-5} \frac{v_\mathrm{esc}}{1 \,\mathrm{m/s}} [\mathrm{m/s^2}]$ at maximum. The escape velocity on Phaethon is $ v_\mathrm{esc} \sim 2\mathrm{-}3 \,\mathrm{m/s} $. As seen in Fig. \ref{fig:acc_diagram-alltifactor1}\textbf{c} and \textbf{d}, $ a_\mathrm{Coriolis, max} $ is still orders of magnitude smaller than any other acceleration terms for Phaethon used in this work, Therefore, Coriolis acceleration is ignored.

For the particles' trajectories, the simplex integration scheme (leapfrog kick-drift-kick) with a fixed time interval of $ dt = 0.025 \,\mathrm{s} $ is used. For selected test cases, longer time intervals (e.g., $ 0.1 \,\mathrm{s} $) sometimes showed artifacts, but time intervals much shorter than $ dt = 0.025 \,\mathrm{s} $  (e.g., $ 0.001 \,\mathrm{s} $) did not change the result. In this preliminary work, Eqs. (\ref{eq: acc_ref}) and (\ref{eq: acc_ther}) are used for all the cases. Nevertheless, it is meaningful to conduct the trajectory integration, because the purpose of this preliminary study is to reveal the {possibility} that small particles can escape from the asteroid.

\begin{figure*}[t!]
  \centering
  \sidecaption
  \includegraphics[width=\linewidth]{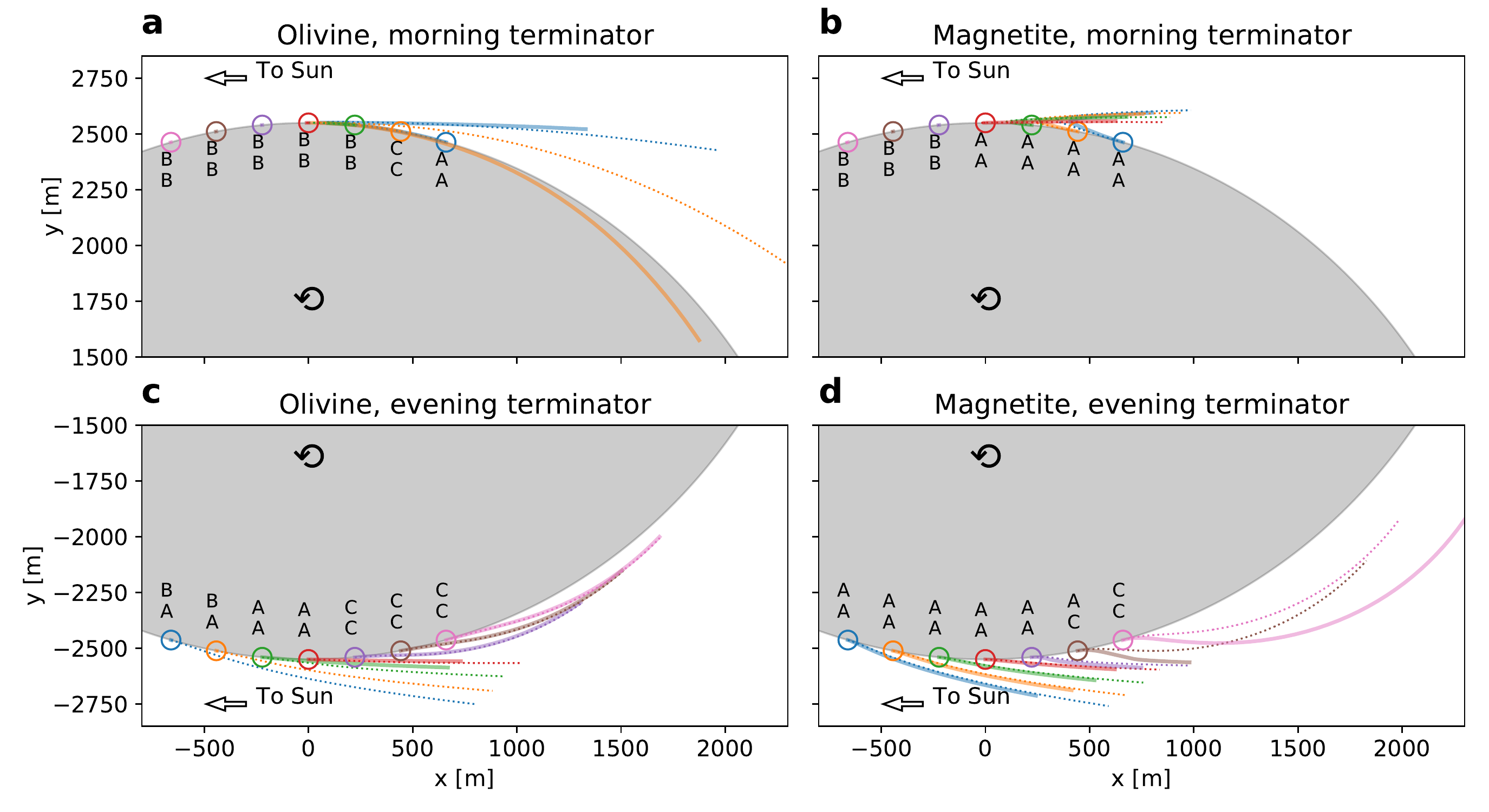}
  \caption{Trajectory of particles projected onto Phaethon's equatorial plane at the perihelion, initially at $ H = 1\cm $ and rotating with Phaethon's rotational speed. Hollow circles, separated by $ 5^\circ $ in longitude, show the initial position of the particles of the same-colored trajectory lines: $ r_a = 1 \um $ (solid) and $ 2 \um $ (dotted). Letters show the flag (see text) of 1 and $ 2 \um $ particles for the first and second line, respectively. The rotation is anti-clockwise as indicated by the arrow, and the Sun is at $ x = -\infty $ as shown by the hallow arrows, but below the $ xy $-plane since the aspect angle is $ 101^\circ > 90^\circ $, so $ a_\mathrm{total} $ has non-zero $ z $ component in this figure. This is why the orange solid line ($ r_a = 1 \um $ at longitude $ 80^\circ $) seems to penetrate into the asteroid, as the trajectory is projected onto the $ xy $-plane. \textbf{a}, \textbf{c}, olivine; \textbf{b}, \textbf{d} magnetite particles. \textbf{a}, \textbf{b}, morning terminator; \textbf{c}, \textbf{d}, evening terminator.}
  \label{fig:phaeacctrajectory}
\end{figure*}

The initial positions of particles are selected by $ \pm 15^\circ $ from the morning and evening terminators at $ 5^\circ $ intervals for $ r_a = 1 $ and $ 2 \um $ (Fig. \ref{fig:phaeacctrajectory}). 

A particle is flagged by three conditions:
\begin{itemize}
\item [(A)] Escape from the gravity field: The radial speed $ \| \mathbf{v} \cdot \rhat \| > v_\mathrm{esc}(r) = \sqrt{2GM/r.} $
\item [(B)] Landing on the surface: The particle height gets smaller than $ 0.1 \cm. $
\item [(C)] Oscillation over the surface: The particle starts to fall back to the surface.
\end{itemize}
The condition (A) is applied when the particles are guaranteed to escape from the gravitational field of Phaethon. In the equation of (A), $ M $ is the Phaethon's mass and $ r $ is the asteroid-centric distance to the particle. $ v_\mathrm{esc} $ on the equator in our model simulation is $ 2.5 \,\mathrm{m/s} $. There was no particle that once reached $ \mathbf{v} \cdot \rhat > v_\mathrm{esc} $ and hit the ground in our test cases. This situation may happen due to the solar radiation when the particle is ejected at high speed towards the Sun. The condition (B) is applied when the particles reach the asteroid surface. The condition (C) captures the artifact caused by our assumptions that underestimate $ a_\mathrm{ther} $ and $ a_\mathrm{ref} $ as the particle height increases. Hence, the particle may fail to further accelerate outwards, even though it could have escaped from the asteroid in reality. There are two ways to flag (C). First, a particle's height is divided by an arbitrary scale and if the quotient (i.e., the integer part of $ \frac{H}{\mathrm{scale}} $) is decreased during the trajectory integration, it is flagged (C). We set $ \mathrm{scale} = 5\,\mathrm{m} $. Another case is when the particle's longitude changes from $ \mathrm{lon} \in [0^\circ, 180^\circ) $ (morning) to $ \mathrm{lon} \in [180^\circ, 360^\circ) $ (evening) or vice versa.

In Fig. \ref{fig:phaeacctrajectory}, the trajectories of particles ($ r_a = 1 $ and $ 2 \um $) are drawn until one of these three flags is applied. The flag is also indicated at the initial position of each particle. If a particle was initially near noon, the trajectory first goes up (increased $ H $) but then down (decreased $ H $) and repeats this oscillatory behavior until it escapes or hits the ground. This is because of the underestimation of $ a_\mathrm{ther} + a_\mathrm{ref} $, while $ a_\odot $ is estimated correctly. This is why the initial positions were chosen only near the terminators in Fig. \ref{fig:phaeacctrajectory}. All the particles with the flag (A) in Fig. \ref{fig:phaeacctrajectory} reach this condition, i.e., escape, within 8 minutes and many reach this condition within 5 minutes, except for olivine initially located at a longitude of $ 75^\circ $, which took 14 and 19 minutes to reach this condition for $ r_a = 1$ and $ 2 \um $, respectively. 

Equation (\ref{eq: acc_odot}) gives $ a_\odot \sim 60\,\mathrm{mm/s^2} $, which is $ \sim 50 a_\mathrm{grav} $, for fiducial particles with $ r_a = 1\,\mathrm{\mu m} $. Thus, ignoring any thermal radiation and assuming $ a_\mathrm{total} \approx a_\odot $ for a first-order approximation (as in \citealt{2013ApJ...771L..36J}), the particle reaches the Hill sphere radius ($r_H = 66\,\mathrm{km}$; \citealt{2019AJ....157..193J}) in 25 minutes or about one-tenth of the rotational period of Phaethon.  Only within about one day will the particle reach the distance of $ 2.5 \times 10^5 \km $, the dust tail length of Phaethon \citep{2013ApJ...771L..36J}. Therefore, the activity should be visible very quickly ($ < 1 \,\mathrm{day}$) after the particles start to be generated.

The fact that most particles could not escape in Fig. \ref{fig:phaeacctrajectory}\textbf{a} implies the importance of thermal radiation: If it were not for $ a_\mathrm{ther} $, particles could not escape easily. For example, let us take a closer look at trajectories of olivine particles in Fig. \ref{fig:phaeacctrajectory}\textbf{a}. Many olivine particles could not escape from Phaethon. 
% This result is expected for all cases regardless of particle composition and size, if it were not for $ a_\mathrm{ther} $, because the solar radiation quickly pushes the particle back into the AR right after the Sunrise (longitude $ \sim 90^\circ $ in Fig. \ref{fig:acc_diagram-alltifactor1}\textbf{c} and \textbf{d}). 
Among them, the olivine particles initially at longitude of $ 75^\circ $ ($ 15^\circ $ before the morning terminator) could nevertheless escape and were flagged as A. What mechanism removed them from the AR? These particles have marginally positive net acceleration, $ \mathbf{a}_\mathrm{total} \cdot \rhat $ ($ a_\mathrm{ther} + a_\mathrm{ref} > a_\mathrm{grav} - a_\mathrm{cen} $ in Fig. \ref{fig:acc_diagram-alltifactor1}). Therefore, due to $ a_\mathrm{ther} $, it is possible for these particles to achieve nonzero outward speeds while rotating above the AR {before}  sunrise. When the Sun rises and $ a_\odot $ comes into play, the particles gain high speed along the anti-solar direction within a few minutes. Then, as the footprints of particles move towards midnight (longitude $ = 0 $ in Fig \ref{fig:acc_diagram-alltifactor1}), the contribution of $ a_\mathrm{ther} $ increases again because of the higher temperature (Fig \ref{fig:acc_diagram-alltifactor1}). Meanwhile, $ v_\mathrm{esc}(r) $ decreases as $ 1/\sqrt{r} $. Eventually, the particles are flagged as (A) at $ x \sim 320 \,\mathrm{m} $ and $ 615 \,\mathrm{m} $ for $ r_a = 1$ and $ 2 \um $, respectively. A similar explanation is applicable for magnetite in Fig. \ref{fig:phaeacctrajectory}\textbf{b}. Magnetite particles in Fig. \ref{fig:phaeacctrajectory}\textbf{b} are more strongly affected by $ a_\mathrm{ther} $ than olivine at a longitude of $ < 90^\circ $ (compare Fig. \ref{fig:acc_diagram-alltifactor1}\textbf{c} and \textbf{d}), and hence escape from Phaethon more easily than olivine. For all particles initially at $ 90^\circ \lesssim \mathrm{longitude} \lesssim 105^\circ $, the net acceleration is always inwards  (toward the AR) in our Phaethon model (Fig. \ref{fig:acc_diagram-alltifactor1}\textbf{c} and \textbf{d}), and therefore they cannot escape and are always flagged as (B).

\begin{figure*}[ht!]
\centering
\sidecaption
\includegraphics[width=\linewidth]{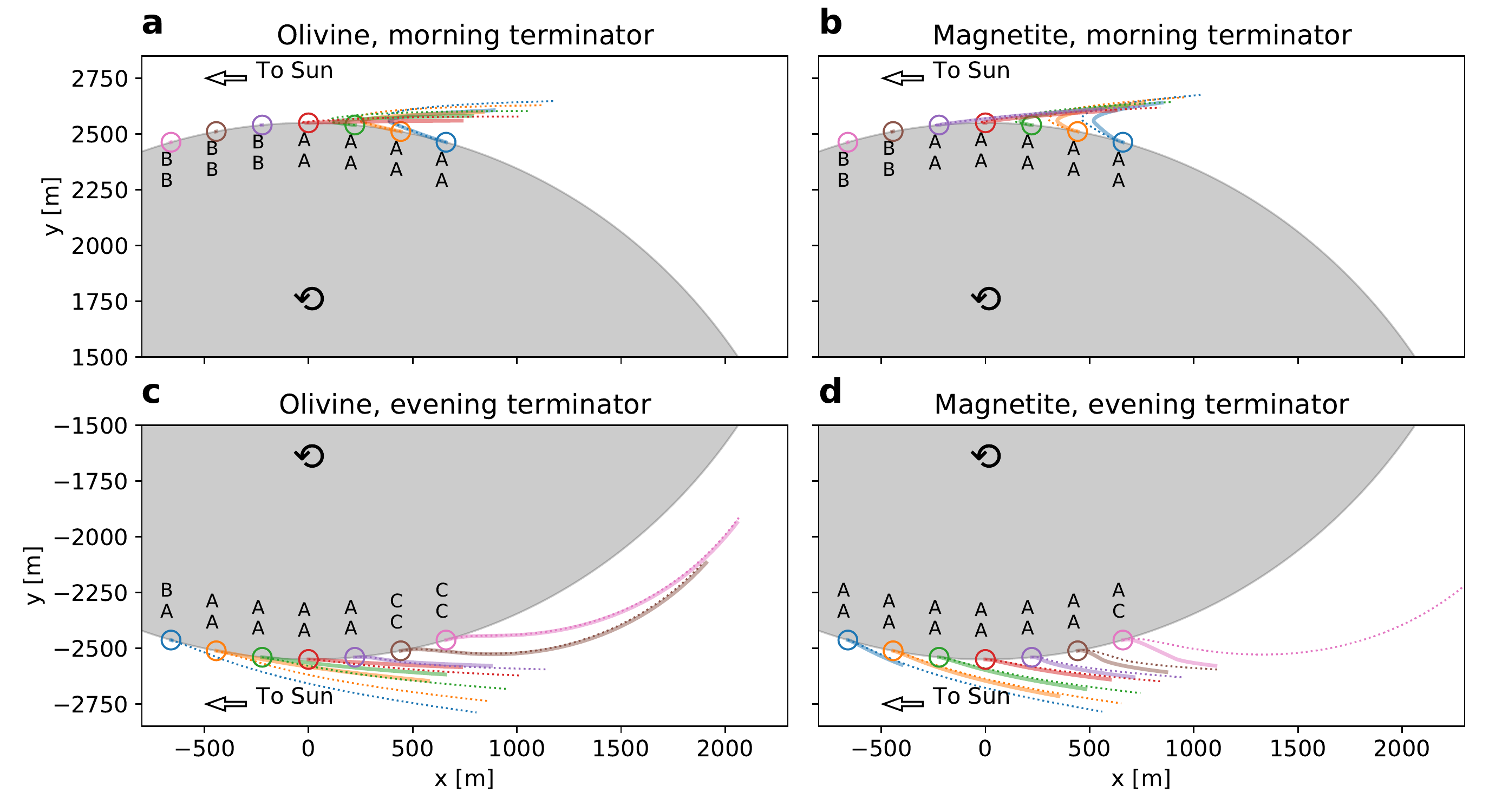}
\caption{Similar figures to Fig. \ref{fig:phaeacctrajectory}, but the thermal inertia $ \Gamma $ is corrected based on Eq. (\ref{eq: TI rh}). We note that many more cases are flagged as (A), i.e., escape-guaranteed. \textbf{a}, \textbf{c}, olivine; \textbf{b}, \textbf{d} magnetite particles. \textbf{a}, \textbf{b}, morning terminator; \textbf{c}, \textbf{d}, evening terminator.}
\label{fig:phaeacctrajectorytirhcorrected}
\end{figure*}

If the thermal inertia $ \Gamma $ is updated as discussed in Sect. \ref{disc-effect of TI}, one would expect particles to become more likely to escape from Phaethon if the initial position is near the terminator. The results are shown in Fig. \ref{fig:phaeacctrajectorytirhcorrected}, and as expected, particles are more easily ejected from the surface. 

\subsection{Phaethon's diameter}
%%%%%%%%%%%%%%%%%%%%%%%%%%%%%%%%%%%%%%%%%%%%%%%%%%%%%%%%%%%%%%%%%%%%%%%%%%%%%%%%
It is important to mention that the diameter of Phaethon is not constrained well among previous publications. In this work, $ D = 5.1 \km $ is adopted, because it is derived by the work that determined the  thermophysical parameters and spin vector in Table \ref{tab: phys_phae} \citep{2016A&A...592A..34H,2018A&A...620L...8H}. A different thermal modeling study suggested $ D = 4.6^{+0.2}_{-0.3} \km $ with geometric albedo $ p_\mathrm{V} = 0.16 \pm 0.02 $ and thermal inertia, $ \Gamma = 880^{+580}_{-330} \tiu $ \citep{2019AJ....158...97M}, whereas a radar study suggested $ D > 6 \km $ \citep{2019P&SS..167....1T}. 

Changes in diameter may affect thermal inertia and/or geometric albedo determination because the albedo is linked with the diameter and the absolute magnitude, and thermal inertia is a parameter used in TPM along with the albedo. It is beyond the scope of this work to discuss the detailed effects of the uncertainties on these parameters. However, the change in the bulk mass density $ \rho_M $ \citep{2018A&A...620L...8H} does not affect our results. $ \rho_M $ is determined by Yarkovsky drift, that is, the change in the semimajor axis of the asteroid. The Yarkovsky acceleration of asteroid $ a_Y $ will be proportional to the reaction force from thermal radiation from the asteroid, $ F_\mathrm{tr} $, divided by its mass, $ M $. Because $ F_\mathrm{tr} \propto D^2 $ and $ M \propto \rho_M D^3 $, the Yarkovsky acceleration $ a_Y \propto 1/\rho_M D $. As $ a_Y $ is the observable, it remains constant, that is, $ \rho_M D = \mathrm{const} $. The gravitational acceleration $ a_\mathrm{grav} \propto \rho_M D $ (Eq. \ref{eq: a_grav}), and therefore it must remain constant even if $ D $ is updated, because $ \rho_M $ is determined from the Yarkovsky drift. Hence, we argue that the update in $ D $ will only lead to a minor correction (e.g., initial speed of particles on the surface) to our results on Phaethon.

\section{Conclusions and future work}
%%%%%%%%%%%%%%%%%%%%%%%%%%%%%%%%%%%%%%%%%%%%%%%%%%%%%%%%%%%%%%%%%%%%%%%%%%%%%%%%
In this work, the importance of thermal radiation from asteroidal regolith surface acting on particles of radii of a few microns is investigated for the first time. Our results show that particles of $ r_a \sim 1\um $ are likely to have the largest outward net acceleration on asteroids in the near-Sun environment (heliocentric distance $ \lesssim 0.8 \au $). The full range of ``ejectable'' particle size spans from $ \sim 0.5 \um $ to $ \lesssim 10 \um $, although the exact range depends on the physical properties of the parent asteroid and particle, as well as initial conditions. This is consistent with observational studies of Phaethon's activity at its perihelion that suggested the ejected particles have $ r_a \sim 1 \um $. 

However, this work should be taken with care because it is agnostic to the mechanisms that generate the particles and achieve the initial height above the regolith, similarly to other works like studies of electrostatic dust levitation. Therefore, future works may include investigations of those mechanisms, and could also include the likelihood of the occurrence of thermal fatigue as functions of heliocentric distance and location on the asteroid and distributions of the generated particles in the size, velocity, and asteroid-centric location domain. Other areas that remain to be investigated are the effects of simplifying assumptions used in thermal modeling. Most importantly, the problem of underestimation of the thermal radiation for heights greater than the isothermal length scale on the asteroid surface must be solved.

\begin{acknowledgements}
This work was supported by the National Research Foundation of Korea (NRF) grant funded by the Korea government (MEST) (No. NRF-2018R1D1A1A09084105, 2019R1A6A1A10073437). Y.P.B. wants to thank Hye-Eun Jang for the discussions on the radiation pressure coefficient and Mie theory, Hojin Cho for providing discussions and priceless comments on many of the details of the manuscript, Harim Jin for discussing the physics of thermal fracturing, Sanghyuk Moon for discussing the numerical calculation of dust particle trajectory, Myung Gyoon Lee for commenting on observational results (all at the same institute as YPB), Scott Prahl at the Oregon Institute of Technology for sharing and revising the Mie theory code, Xu Wang at the University of Colorado, Boulder, and Xiao Ping Zhang at Macau University of Science and Technology for their kind discussions and explanations about the electrostatic dust lofting mechanism, Josef Hanu\v{s} at Charles University in Prague for sharing information on how the widely used thermal modeling code works, and Harald Mutschke at University of Jena for kind discussions on the optical index experiments of olivine and magnetite. We thank the reviewer, Oleksiy Golubov, for thoroughly reviewing the manuscript and greatly improved its quality in all aspects.
\end{acknowledgements}

\bibliographystyle{aa}
\bibliography{thermal}

\begin{thebibliography}{52}
\expandafter\ifx\csname natexlab\endcsname\relax\def\natexlab#1{#1}\fi

\bibitem[{{Arai} {et~al.}(2018){Arai}, {Kobayashi}, {Ishibashi}, {Yoshida},
  {Kimura}, {Wada}, {Senshu}, {Yamada}, {Okudaira}, {Okamoto}, {Kameda},
  {Srama}, {Kruger}, {Ishiguro}, {Yabuta}, {Nakamura}, {Watanabe}, {Ito},
  {Ohtsuka}, {Tachibana}, {Mikouchi}, {Komatsu}, {Nakamura-Messenger},
  {Sasaki}, {Hiroi}, {Abe}, {Urakawa}, {Hirata}, {Demura}, {Komatsu},
  {Noguchi}, {Sekiguchi}, {Inamori}, {Yano}, {Yoshikawa}, {Ohtsubo}, {Okada},
  {Iwata}, {Nishiyama}, {Toyota}, {Kawakatsu}, \&
  {Takashima}}]{2018LPI....49.2570A}
{Arai}, T., {Kobayashi}, M., {Ishibashi}, K., {et~al.} 2018, in Lunar and
  Planetary Science Conference, Lunar and Planetary Science Conference, 2570

\bibitem[{{Blaauw}(2017)}]{2017P&SS..143...83B}
{Blaauw}, R.~C. 2017, \planss, 143, 83

\bibitem[{{Bohren} \& {Huffman}(1983)}]{1983asls.book.....B}
{Bohren}, C.~F. \& {Huffman}, D.~R. 1983, {Absorption and scattering of light
  by small particles}

\bibitem[{{Bohren} \& {Huffman}(1998)}]{1998asls.book.....B}
{Bohren}, C.~F. \& {Huffman}, D.~R. 1998, {Absorption and Scattering of Light
  by Small Particles}

\bibitem[{{Bowell} {et~al.}(1989){Bowell}, {Hapke}, {Domingue}, {Lumme},
  {Peltoniemi}, \& {Harris}}]{1989aste.conf..524B}
{Bowell}, E., {Hapke}, B., {Domingue}, D., {et~al.} 1989, in Asteroids II, ed.
  R.~P. {Binzel}, T.~{Gehrels}, \& M.~S. {Matthews}, 524--556

\bibitem[{{Carry}(2012)}]{2012P&SS...73...98C}
{Carry}, B. 2012, \planss, 73, 98

\bibitem[{{Chamberlin} {et~al.}(1996){Chamberlin}, {McFadden}, {Schulz},
  {Schleicher}, \& {Bus}}]{1996Icar..119..173C}
{Chamberlin}, A.~B., {McFadden}, L.-A., {Schulz}, R., {Schleicher}, D.~G., \&
  {Bus}, S.~J. 1996, \icarus, 119, 173

\bibitem[{Delbo(2004)}]{DelboM2004PhDT}
Delbo, M. 2004, PhD thesis, Free University of Berlin

\bibitem[{{Delbo'} {et~al.}(2007){Delbo'}, {dell'Oro}, {Harris}, {Mottola}, \&
  {Mueller}}]{2007Icar..190..236D}
{Delbo'}, M., {dell'Oro}, A., {Harris}, A.~W., {Mottola}, S., \& {Mueller}, M.
  2007, \icarus, 190, 236

\bibitem[{{Delbo} {et~al.}(2014){Delbo}, {Libourel}, {Wilkerson}, {Murdoch},
  {Michel}, {Ramesh}, {Ganino}, {Verati}, \& {Marchi}}]{2014Natur.508..233D}
{Delbo}, M., {Libourel}, G., {Wilkerson}, J., {et~al.} 2014, \nat, 508, 233

\bibitem[{{Delbo} {et~al.}(2015){Delbo}, {Mueller}, {Emery}, {Rozitis}, \&
  {Capria}}]{2015aste.book..107D}
{Delbo}, M., {Mueller}, M., {Emery}, J.~P., {Rozitis}, B., \& {Capria}, M.~T.
  2015, {Asteroid Thermophysical Modeling}, 107--128

\bibitem[{{Fabian} {et~al.}(2001){Fabian}, {Henning}, {J{\"a}ger}, {Mutschke},
  {Dorschner}, \& {Wehrhan}}]{2001A&A...378..228F}
{Fabian}, D., {Henning}, T., {J{\"a}ger}, C., {et~al.} 2001, \aap, 378, 228

\bibitem[{{Green} {et~al.}(1985){Green}, {Meadows}, \&
  {Davies}}]{1985MNRAS.214P..29G}
{Green}, S.~F., {Meadows}, A.~J., \& {Davies}, J.~K. 1985, \mnras, 214, 29P

\bibitem[{{Grott} {et~al.}(2019){Grott}, {Knollenberg}, {Hamm}, {Ogawa},
  {Jaumann}, {Otto}, {Delbo}, {Michel}, {Biele}, {Neumann}, {Knapmeyer},
  {K{\"u}hrt}, {Senshu}, {Okada}, {Helbert}, {Maturilli}, {M{\"u}ller},
  {Hagermann}, {Sakatani}, {Tanaka}, {Arai}, {Mottola}, {Tachibana}, {Pelivan},
  {Drube}, {Vincent}, {Yano}, {Pilorget}, {Matz}, {Schmitz}, {Koncz},
  {Schr{\"o}der}, {Trauthan}, {Schlotterer}, {Krause}, {Ho}, \&
  {Moussi-Soffys}}]{2019NatAs...3..971G}
{Grott}, M., {Knollenberg}, J., {Hamm}, M., {et~al.} 2019, Nature Astronomy, 3,
  971

\bibitem[{{Hanu{\v{s}}} {et~al.}(2016){Hanu{\v{s}}}, {Delbo'},
  {Vokrouhlick{\'y}}, {Pravec}, {Emery}, {Al{\'\i}-Lagoa}, {Bolin},
  {Devog{\`e}le}, {Dyvig}, {Gal{\'a}d}, {Jedicke}, {Korno{\v{s}}},
  {Ku{\v{s}}nir{\'a}k}, {Licandro}, {Reddy}, {Rivet}, {Vil{\'a}gi}, \&
  {Warner}}]{2016A&A...592A..34H}
{Hanu{\v{s}}}, J., {Delbo'}, M., {Vokrouhlick{\'y}}, D., {et~al.} 2016, \aap,
  592, A34

\bibitem[{{Hanu{\v{s}}} {et~al.}(2018){Hanu{\v{s}}}, {Vokrouhlick{\'y}},
  {Delbo'}, {Farnocchia}, {Polishook}, {Pravec}, {Hornoch},
  {Ku{\v{c}}{\'a}kov{\'a}}, {Ku{\v{s}}nir{\'a}k}, {Stephens}, \&
  {Warner}}]{2018A&A...620L...8H}
{Hanu{\v{s}}}, J., {Vokrouhlick{\'y}}, D., {Delbo'}, M., {et~al.} 2018, \aap,
  620, L8

\bibitem[{Hapke(2012)}]{hapke2012book}
Hapke, B. 2012, Theory of reflectance and emittance spectroscopy (Cambridge
  university press)

\bibitem[{{Harris}(1998)}]{1998Icar..131..291H}
{Harris}, A.~W. 1998, \icarus, 131, 291

\bibitem[{{Hartzell} \& {Scheeres}(2011)}]{2011P&SS...59.1758H}
{Hartzell}, C.~M. \& {Scheeres}, D.~J. 2011, \planss, 59, 1758

\bibitem[{{Hesar} {et~al.}(2017){Hesar}, {Scheeres}, {McMahon}, \&
  {Rozitis}}]{2017JGCD...40.2432H}
{Hesar}, S.~G., {Scheeres}, D.~J., {McMahon}, J.~W., \& {Rozitis}, B. 2017,
  Journal of Guidance Control Dynamics, 40, 2432

\bibitem[{{Hood} {et~al.}(2018){Hood}, {Carroll}, {Mike}, {Wang}, {Schwan},
  {Hsu}, \& {Hor{\'a}nyi}}]{2018GeoRL..4513206H}
{Hood}, N., {Carroll}, A., {Mike}, R., {et~al.} 2018, \grl, 45, 13,206

\bibitem[{{Hsieh} \& {Jewitt}(2005)}]{2005ApJ...624.1093H}
{Hsieh}, H.~H. \& {Jewitt}, D. 2005, \apj, 624, 1093

\bibitem[{{Hui} \& {Li}(2017)}]{2017AJ....153...23H}
{Hui}, M.-T. \& {Li}, J. 2017, \aj, 153, 23

\bibitem[{{Jewitt}(2012)}]{2012AJ....143...66J}
{Jewitt}, D. 2012, \aj, 143, 66

\bibitem[{{Jewitt} {et~al.}(2019){Jewitt}, {Asmus}, {Yang}, \&
  {Li}}]{2019AJ....157..193J}
{Jewitt}, D., {Asmus}, D., {Yang}, B., \& {Li}, J. 2019, \aj, 157, 193

\bibitem[{{Jewitt} \& {Li}(2010)}]{2010AJ....140.1519J}
{Jewitt}, D. \& {Li}, J. 2010, \aj, 140, 1519

\bibitem[{{Jewitt} {et~al.}(2013){Jewitt}, {Li}, \&
  {Agarwal}}]{2013ApJ...771L..36J}
{Jewitt}, D., {Li}, J., \& {Agarwal}, J. 2013, \apjl, 771, L36

\bibitem[{{Jewitt} {et~al.}(2018){Jewitt}, {Mutchler}, {Agarwal}, \&
  {Li}}]{2018AJ....156..238J}
{Jewitt}, D., {Mutchler}, M., {Agarwal}, J., \& {Li}, J. 2018, \aj, 156, 238

\bibitem[{{Lauretta} {et~al.}(2019){Lauretta}, {Hergenrother}, {Chesley},
  {Leonard}, {Pelgrift}, {Adam}, {Al Asad}, {Antreasian}, {Ballouz}, {Becker},
  {Bennett}, {Bos}, {Bottke}, {Brozovi{\'c}}, {Campins}, {Connolly}, {Daly},
  {Davis}, {de Le{\'o}n}, {DellaGiustina}, {Drouet d'Aubigny}, {Dworkin},
  {Emery}, {Farnocchia}, {Glavin}, {Golish}, {Hartzell}, {Jacobson}, {Jawin},
  {Jenniskens}, {Kidd}, {Lessac-Chenen}, {Li}, {Libourel}, {Licandro},
  {Liounis}, {Maleszewski}, {Manzoni}, {May}, {McCarthy}, {McMahon}, {Michel},
  {Molaro}, {Moreau}, {Nelson}, {Owen}, {Rizk}, {Roper}, {Rozitis}, {Sahr},
  {Scheeres}, {Seabrook}, {Selznick}, {Takahashi}, {Thuillet}, {Tricarico},
  {Vokrouhlick{\'y}}, \& {Wolner}}]{2019Sci...366.3544L}
{Lauretta}, D.~S., {Hergenrother}, C.~W., {Chesley}, S.~R., {et~al.} 2019,
  Science, 366, 3544

\bibitem[{{Li} \& {Jewitt}(2013)}]{2013AJ....145..154L}
{Li}, J. \& {Jewitt}, D. 2013, \aj, 145, 154

\bibitem[{{Luu} \& {Jewitt}(1992)}]{1992Icar...97..276L}
{Luu}, J.~X. \& {Jewitt}, D.~C. 1992, \icarus, 97, 276

\bibitem[{{Masiero} {et~al.}(2019){Masiero}, {Wright}, \&
  {Mainzer}}]{2019AJ....158...97M}
{Masiero}, J.~R., {Wright}, E.~L., \& {Mainzer}, A.~K. 2019, \aj, 158, 97

\bibitem[{McMahon {et~al.}(2020)McMahon, Scheeres, Chesley, French, Brack,
  Farnocchia, Takahashi, Rozitis, Tricarico, Mazarico, Bierhaus, Emery,
  Hergenrother, \& Lauretta}]{McMahon2020}
McMahon, J.~W., Scheeres, D.~J., Chesley, S.~R., {et~al.} 2020, Journal of
  Geophysical Research: Planets

\bibitem[{{Molaro} {et~al.}(2017){Molaro}, {Byrne}, \&
  {Le}}]{2017Icar..294..247M}
{Molaro}, J.~L., {Byrne}, S., \& {Le}, J.~L. 2017, \icarus, 294, 247

\bibitem[{Mueller(2007)}]{MuellerM2007PhDT}
Mueller, M. 2007, PhD thesis, Free University of Berlin

\bibitem[{{Myhrvold}(2016)}]{2016PASP..128d5004M}
{Myhrvold}, N. 2016, \pasp, 128, 045004

\bibitem[{{Orger} {et~al.}(2018){Orger}, {Cordova Alarcon}, {Toyoda}, \&
  {Cho}}]{2018AdSpR..62..896O}
{Orger}, N.~C., {Cordova Alarcon}, J.~R., {Toyoda}, K., \& {Cho}, M. 2018,
  Advances in Space Research, 62, 896

\bibitem[{{Putzig}(2006)}]{2006PhDT........15P}
{Putzig}, N.~E. 2006, PhD thesis, University of Colorado at Boulder

\bibitem[{{Ravaji} {et~al.}(2019){Ravaji}, {Al{\'\i}-Lagoa}, {Delbo}, \&
  {Wilkerson}}]{2019JGRE..124.3304R}
{Ravaji}, B., {Al{\'\i}-Lagoa}, V., {Delbo}, M., \& {Wilkerson}, J.~W. 2019,
  Journal of Geophysical Research (Planets), 124, 3304

\bibitem[{{Righter} {et~al.}(2006){Righter}, {Drake}, \&
  {Scott}}]{2006mess.book..803R}
{Righter}, K., {Drake}, M.~J., \& {Scott}, E.~R.~D. 2006, {Compositional
  Relationships Between Meteorites and Terrestrial Planets}, ed. D.~S.
  {Lauretta} \& H.~Y. {McSween}, 803

\bibitem[{{Robie} \& {Bethke}(1962)}]{robie1962}
{Robie}, R.~A. \& {Bethke}, P.~M. 1962, Molar volumes and densities of minerals
  (Unites States Department of the Interior Geological Survey)

\bibitem[{{Ryabova}(2017)}]{2017P&SS..143..125R}
{Ryabova}, G.~O. 2017, \planss, 143, 125

\bibitem[{{Scott} \& {Krot}(2005)}]{2005ApJ...623..571S}
{Scott}, E. R.~D. \& {Krot}, A.~N. 2005, \apj, 623, 571

\bibitem[{{Senshu} {et~al.}(2015){Senshu}, {Kimura}, {Yamamoto}, {Wada},
  {Kobayashi}, {Namiki}, \& {Matsui}}]{2015P&SS..116...18S}
{Senshu}, H., {Kimura}, H., {Yamamoto}, T., {et~al.} 2015, \planss, 116, 18

\bibitem[{{Spencer} {et~al.}(1989){Spencer}, {Lebofsky}, \&
  {Sykes}}]{1989Icar...78..337S}
{Spencer}, J.~R., {Lebofsky}, L.~A., \& {Sykes}, M.~V. 1989, \icarus, 78, 337

\bibitem[{{Taylor} {et~al.}(2019){Taylor}, {Rivera-Valent{\'\i}n}, {Benner},
  {Marshall}, {Virkki}, {Venditti}, {Zambrano-Marin}, {Bhiravarasu},
  {Aponte-Hernandez}, {Rodriguez Sanchez-Vahamonde}, \&
  {Giorgini}}]{2019P&SS..167....1T}
{Taylor}, P.~A., {Rivera-Valent{\'\i}n}, E.~G., {Benner}, L. A.~M., {et~al.}
  2019, \planss, 167, 1

\bibitem[{{Tsuchiyama} {et~al.}(2011){Tsuchiyama}, {Uesugi}, {Matsushima},
  {Michikami}, {Kadono}, {Nakamura}, {Uesugi}, {Nakano}, {Sandford}, {Noguchi},
  {Matsumoto}, {Matsuno}, {Nagano}, {Imai}, {Takeuchi}, {Suzuki}, {Ogami},
  {Katagiri}, {Ebihara}, {Ireland}, {Kitajima}, {Nagao}, {Naraoka}, {Noguchi},
  {Okazaki}, {Yurimoto}, {Zolensky}, {Mukai}, {Abe}, {Yada}, {Fujimura},
  {Yoshikawa}, \& {Kawaguchi}}]{2011Sci...333.1125T}
{Tsuchiyama}, A., {Uesugi}, M., {Matsushima}, T., {et~al.} 2011, Science, 333,
  1125

\bibitem[{{van de Hulst}(1981)}]{1981lssp.book.....V}
{van de Hulst}, H.~C. 1981, {Light scattering by small particles}

\bibitem[{{Wang} {et~al.}(2016){Wang}, {Schwan}, {Hsu}, {Gr{\"u}n}, \&
  {Hor{\'a}nyi}}]{2016GeoRL..43.6103W}
{Wang}, X., {Schwan}, J., {Hsu}, H.~W., {Gr{\"u}n}, E., \& {Hor{\'a}nyi}, M.
  2016, \grl, 43, 6103

\bibitem[{{Warner} {et~al.}(2009){Warner}, {Harris}, \&
  {Pravec}}]{2009Icar..202..134W}
{Warner}, B.~D., {Harris}, A.~W., \& {Pravec}, P. 2009, \icarus, 202, 134

\bibitem[{{Wiegert} {et~al.}(2008){Wiegert}, {Houde}, \&
  {Peng}}]{2008Icar..194..843W}
{Wiegert}, P.~A., {Houde}, M., \& {Peng}, R. 2008, \icarus, 194, 843

\bibitem[{{Zimmerman} {et~al.}(2016){Zimmerman}, {Farrell}, {Hartzell}, {Wang},
  {Horanyi}, {Hurley}, \& {Hibbitts}}]{2016JGRE..121.2150Z}
{Zimmerman}, M.~I., {Farrell}, W.~M., {Hartzell}, C.~M., {et~al.} 2016, Journal
  of Geophysical Research (Planets), 121, 2150

\end{thebibliography}

\begin{appendix}

\section{Radiation pressure coefficients ($ Q_\mathrm{pr} (\lambda, r_a) $)} \label{app: qpr}
\begin{figure*}
  \centering
  \includegraphics[width=\linewidth]{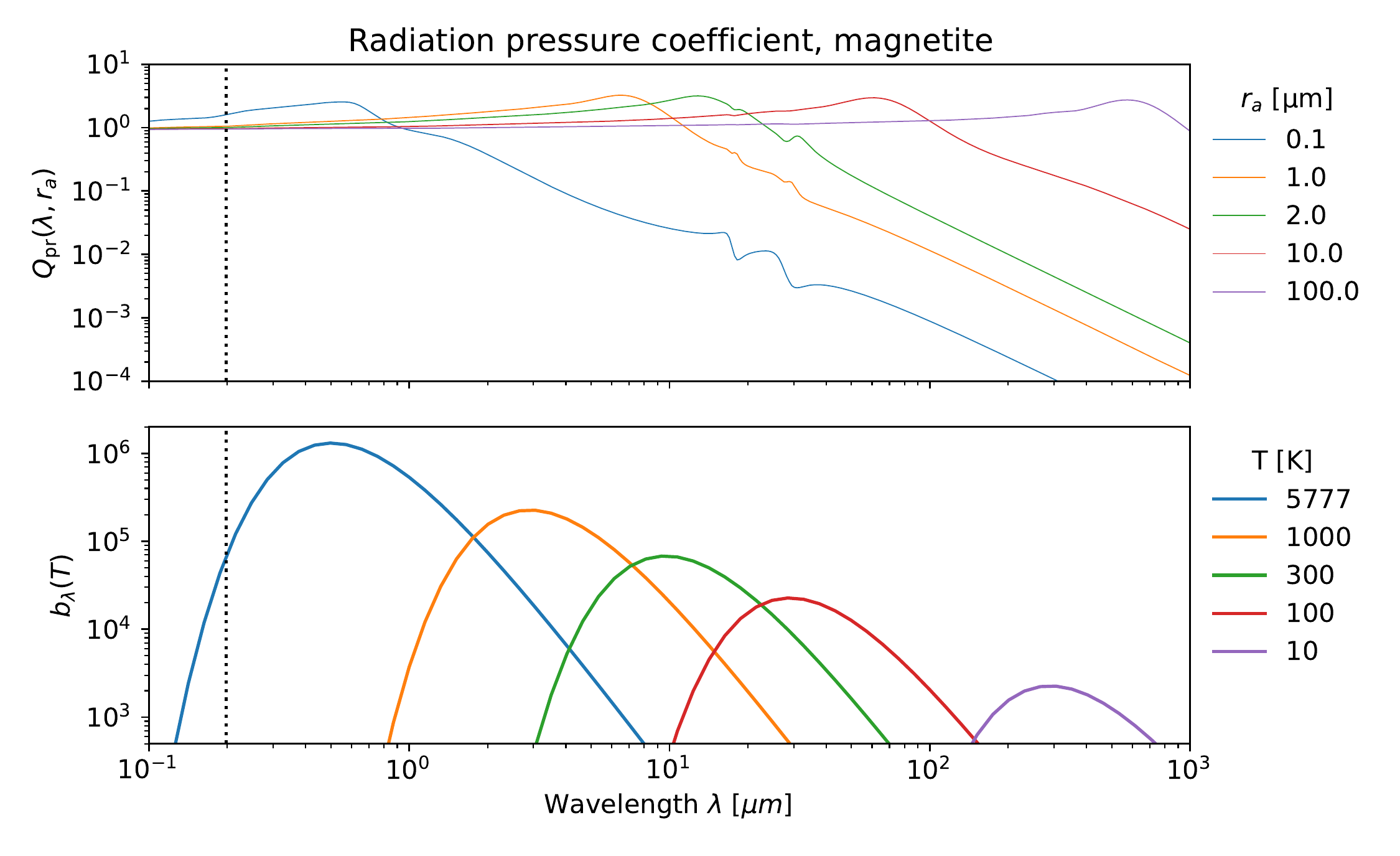}
  \caption{Radiation pressure coefficient, $ Q_\mathrm{pr}(\lambda, r_a) $ of magnetite (top) and the function $ b_\lambda(\lambda, T) $ (bottom; Eq. \ref{eq: b lambda}) for selected temperatures.}
  \label{fig:qprmag}
\end{figure*}

\begin{figure*}
  \centering
  \includegraphics[width=\linewidth]{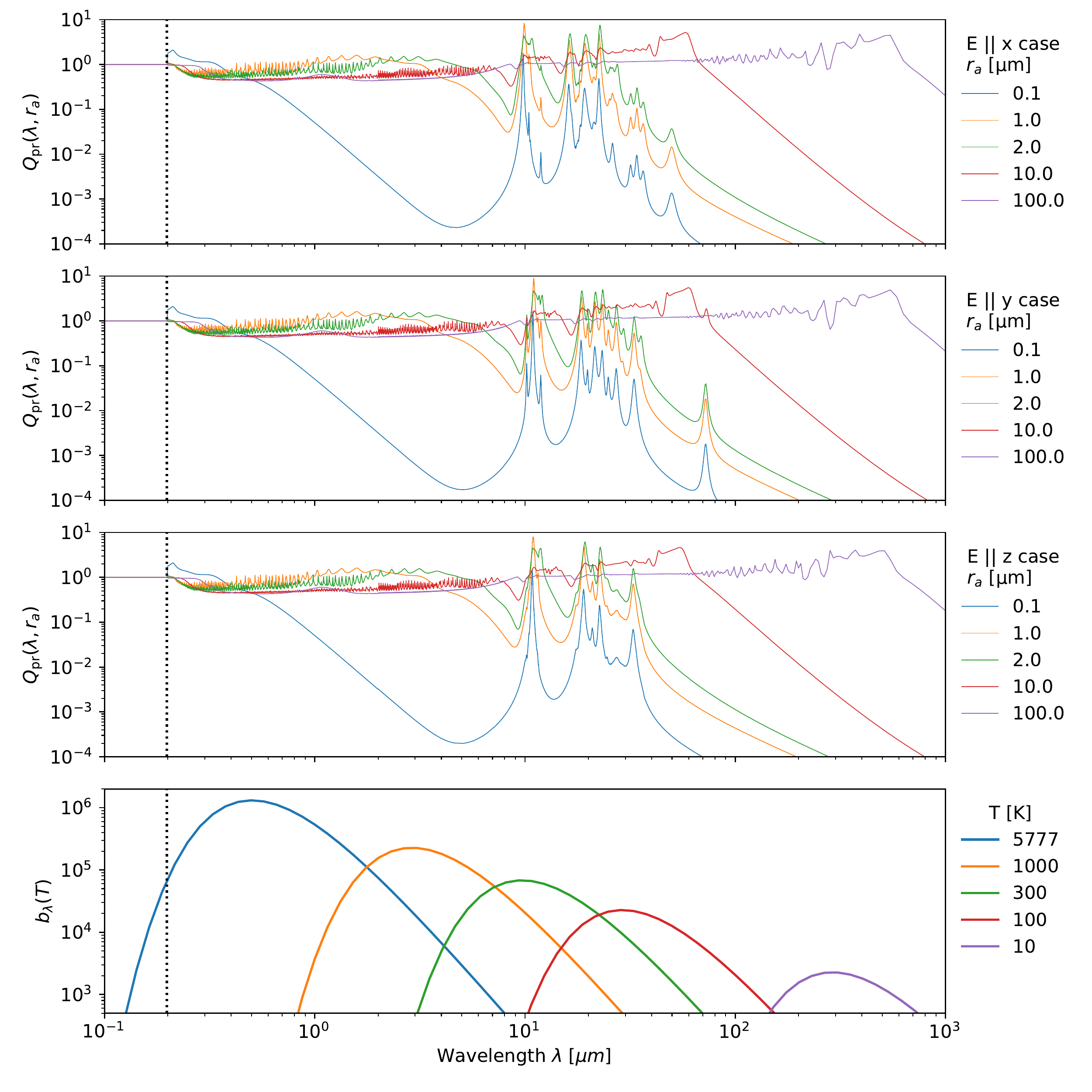}
  \caption{Radiation pressure coefficient, $ Q_\mathrm{pr}(\lambda, r_a) $ of olivine (first three rows) and the function $ b_\lambda(\lambda, T) $ (bottom; Eq. \ref{eq: b lambda}) for selected temperatures as in Fig. \ref{fig:qprmag}. The three different $ Q_\mathrm{pr} $ values are derived from three different optical indices depending on the polarization of the incident light with respect to the crystalline structure (see text).}
  \label{fig:qproliv}
\end{figure*}

As described in Sect. \ref{meth-mie}, the radiation pressure coefficient, $ Q_\mathrm{pr}(\lambda, r_a) $, is calculated for two materials, olivine and magnetite. The calculated $ Q_\mathrm{pr} $ values are shown as functions of wavelength $ \lambda $ for selected particle radii in Fig. \ref{fig:qproliv} (olivine) and Fig. \ref{fig:qprmag} (magnetite). 

The optical index is available at $ \lambda > 0.1 \um $ for magnetite (Sect. \ref{meth-mie}) The derived $ Q_\mathrm{pr} $ from Mie theory is shown in Fig. \ref{fig:qprmag}.

For (low-Fe) olivine, two datasets of optical indices are combined (\citealt{2001A&A...378..228F}; Sect. \ref{meth-mie}).  The first dataset is composed of three indices as functions of wavelength $ \lambda = 2.0\mathrm{-}4097 \um$. The three indices are obtained from incident light polarized parallel to each of the three different crystalline axes of olivine (orthorhombic). These are denoted $ E \| x $, $ E \| y $, or $ E \| z $ following the notation in the original publication. The second dataset is obtained based on a measurement for a randomly oriented olivine sample in $ \lambda = 0.2 \mathrm{-} 2.0 \um $. After combining the two datasets, the optical index is identical for $ \lambda < 2.0 \um $, but there are three separate cases for $ \lambda > 2.0 \um $. These three different optical indices result in three separate $ Q_\mathrm{pr} $ curves in Fig. \ref{fig:qproliv} and $ \bar{Q}_\mathrm{pr} $ curves in Fig \ref{fig:qprbar}. 

We further assumed $ Q_\mathrm{pr} = 1 $ for $ 0.1 \um \le \lambda \le 0.2 \um $ for olivine to match the wavelength range with magnetite. For small particles of $ r_a \ll 1 \um $, the value of $ Q_\mathrm{pr} (\lambda \in [0.1,\, 0.2] \um, r_a) $ may be different from unity (as in the $ r_a = 0.1 \um $ curve in Fig. \ref{fig:qproliv}). However, this deviation is not critical because the contribution of the radiation flux at the short wavelengths to the total acceleration is small (for blackbody radiation of temperature $ T \lesssim T_\odot $). Furthermore, as described in Sect. \ref{meth-radacc} and Fig. \ref{fig:qprbar}, the resulting difference in $ \bar{Q}_\mathrm{pr} $ is negligible. Hence, $ \bar{Q}_\mathrm{pr} $ for $ E \| x $, $ E \| y $, and $ E \| z $ are averaged to get olivine's $ \bar{Q}_\mathrm{pr} $ value used throughout this study.

\section{Radiative accelerations derivations} \label{app: acc}
In this section, we derive the equations we used in Sect. \ref{meth-radacc}. 

\subsection{Auxiliary notations} \label{app: acc notation}
The blackbody spectrum-averaged emissivity can be defined as
\begin{equation} \label{eq: epsilon-bar def}
  \bar{\epsilon} (T, \epsilon)
    \equiv \frac{\int_{0}^{\infty} \epsilon(\lambda) B_\lambda(\lambda, T) d\lambda }
      {\int_{0}^{\infty} B_\lambda(\lambda, T) d\lambda } ~.
\end{equation}
We derived Eq. (\ref{eq: epsilon-bar}) using the Stefan--Boltzmann law,
\begin{equation} \label{eq: stefan-boltzmann}
  \int_{0}^{\infty} B_\lambda(\lambda, T) d\lambda = \frac{\sigma_\mathrm{SB} T^4}{\pi} ~.
\end{equation}
 Next, the radiance (Planck function multiplied by the emissivity) can be normalized by the total flux:
\begin{equation}\label{eq: b lambda def}
  b_\lambda (\lambda, T, \epsilon) 
    \equiv \frac{\epsilon(\lambda) B_\lambda(\lambda, T) d\lambda }
    {\int_{0}^{\infty} \epsilon(\lambda) B_\lambda(\lambda, T) d\lambda}.
\end{equation}
Subsequently, substituting Eq. (\ref{eq: epsilon-bar def}) and (\ref{eq: stefan-boltzmann}), we derive Eq. (\ref{eq: b lambda}). Similarly, $ Q_\mathrm{pr} $ can be averaged over the radiance, i.e., 
\begin{equation} \label{eq: bar Qpr def}
  \bar{Q}_\mathrm{pr}(r_a, T, \epsilon) 
    \equiv \frac{\int_{0}^{\infty} Q_\mathrm{pr}(\lambda, r_a) \epsilon(\lambda) B_\lambda(\lambda, T) d\lambda}
    {\int_{0}^{\infty} \epsilon(\lambda) B_\lambda(\lambda, T) d\lambda}
~.
\end{equation}
We then use  Eq. (\ref{eq: b lambda}) to derive Eq. (\ref{eq: bar Qpr}). Using the above notations, a simplified notation without integral is developed:
\begin{equation}\label{eq: int_QeB}
  \int_{0}^{\infty} Q_\mathrm{pr}(\lambda, r_a) \epsilon(\lambda) B_\lambda(\lambda, T) d\lambda
  = \frac{\sigma_\mathrm{SB} T^4}{\pi} \bar{Q}_\mathrm{pr} \bar{\epsilon} ~.
\end{equation}
This relation is used to derive Eq. (\ref{eq: a_rad}).

\subsection{Derivation of Eq. (\ref{eq: a_rad})} \label{app: acc acc}
When an object with a geometric cross-sectional area of $ \sigma = \pi r_a^2 $ is illuminated by a spectral irradiance of $ J_\lambda (\lambda) $, the magnitude of the radiative force, i.e., the momentum flux per time, is
\begin{equation}
  \| \mathbf{F} \| = \frac{1}{c} \int_{0}^{\infty} J_\lambda(\lambda) \sigma Q_\mathrm{pr}(\lambda, r_a) d\lambda ~.
\end{equation}
The magnitude of the radiative acceleration that the particle experiences is $ a = \| \mathbf{F} \|/m $, where $ m = 4\pi \rho_m r_a^3/3 $ is  the particle mass. If the source is a blackbody with a solid angle of $ d\Omega $ and emissivity $ \epsilon $, the irradiance $ J_\lambda = \epsilon B_\lambda d \Omega $. Substituting $ m $, $ J_\lambda $, and Eq. (\ref{eq: int_QeB}), $ a $ can be rewritten as
\begin{equation}\label{eq: a_rad def}
  a = \frac{3\sigma_\mathrm{SB}}{4c} \frac{\bar{\epsilon} \bar{Q}_\mathrm{pr}(r_a, T)}{r_a \rho_m} T^4 \frac{d \Omega}{\pi} ~,
\end{equation}
which is Eq. (\ref{eq: a_rad}).

\subsection{Derivation of Eq. (\ref{eq: acc_ref})} \label{app: acc_ref}
For the $ a_\mathrm{ref} $ calculation, the irradiance must be integrated over the surface elements within the field of view. As we assume the AR is a Lambertian scatterer, it leads to
\begin{equation}
r_\mathrm{ref}(i) = \frac{A_B}{\pi} \cos i ~,
\end{equation}
where $ r_\mathrm{ref} = r_\mathrm{ref}(\lambda, \hat{\Omega}_i, \hat{\Omega}_{e'}) $ is the bidirectional reflectance, and $ A_B $ is the Bond albedo of the Lambertian patch \cite[for a derivation, see, e.g.,][Sect. 8.5.1]{hapke2012book}. 
% lengthy comment on this (removed)
%If a convex asteroid has identical materials on the surface, i.e., the asteroid has no shadowing and $ A_B = \mathrm{const} $, the disk-integrated (spherical) Bond albedo is identical to the (hemispherical) Bond albedo of each surface element. Therefore, we can adopt any Bond albedo of a realistic asteroid as the Bond albedo of the AR by assuming the asteroid is a Lambertian scatterer.
Therefore, the irradiance from the infinitesimal patch of solid angle $ d\Omega $ is 
\begin{equation}\label{eq: dJ_ref}
  dJ_\mathrm{\lambda, ref} 
    = J_{\lambda, \odot} r_\mathrm{ref} d\Omega 
    = J_{\lambda, \odot} A_B \mu_i \frac{d\Omega}{\pi} ~,
\end{equation}
where $ \mu_i = \mathrm{max} \, \{\cos i,\, 0\} $. By using $ \mu_i $, the reflected radiation is ``turned off'' immediately after the sunset when calculated by a code. 

The Lambertian scattering leads to an azimuthal symmetry of the scattered radiation from the local surface, so the acceleration of the particle perpendicular to the $ z $-axis will cancel out. The acceleration along the $ z $-axis from the infinitesimal patch is therefore (see Fig. \ref{fig:fschem_ref_th}) 
\begin{equation}\label{eq: da_ref}
  da_\mathrm{ref} = A_B \mu_i \frac{d\Omega}{\pi} \cos e'' \times a_\odot ~.
\end{equation}
The infinitesimal \textit{annulus} at angle $ e'' $ and width $ de'' $ has the solid angle of $ d\Omega_{e''} = 2\pi \sin e'' de'' $, and any patch within this annulus will give an identical acceleration contribution to the particle along the $ z $-direction. Thus, the contribution of the annulus is obtained by replacing $ d\Omega $ with $ d\Omega_{e''} $ in Eq (\ref{eq: da_ref}). The integration of the annulus of radius from $ 0 $ to $ r_0 $ is identical to the integration of $ e'' $ from $ 0 $ to $ e''_\mathrm{max} = \tan^{-1}(r_0/H) $. If the height factor $ \tilde{H} $ is defined as Eq. (\ref{eq: H factor}), 
\begin{equation}
  \int_{0}^{e''_\mathrm{max}} \cos e'' \sin e'' de'' = \frac{\tilde{H}}{2} ~.
\end{equation}
Thus, after performing the integration, Eq. (\ref{eq: acc_ref}) is obtained.

\subsection{Derivation of Eq. (\ref{eq: acc_ther})} \label{app: acc_ther}
For $ a_\mathrm{ther} $, similar to $ a_\mathrm{ref} $, the azimuthal symmetry is guaranteed. Meanwhile, the acceleration ($ da_\mathrm{ther} $) due to the radiation originating from an infinitesimal area on the AR with solid angle $ d\Omega_{e''} $ is given by Eq. (\ref{eq: a_rad def}), with $ \epsilon = \bar{\epsilon}_S $, $ T = T_S $, and $ d\Omega = d\Omega_{e''} $. Then, multiplying $ \cos e'' $ and integrating over $ d\Omega_{e''} $, in the same way as done for $ a_\mathrm{ref} $, leads to Eq. (\ref{eq: acc_ther}).

\section{Thermal fatigue reparameterization} \label{app: fatigue reparam}
Here, we derive the two criteria we used in Sect. \ref{disc-thermal fatigue}. The two criteria are from \cite{2019JGRE..124.3304R}, but we devise a more general view using the $ (\rh,\, \Theta) $ coordinate by re-formulating them. 

For the first one, $ \Delta \hat{T}_S > 100\,\mathrm{K} $, we rewrite 
\begin{equation*}
  \Delta \hat{T}_S 
    = \frac{\Delta \hat{T}_S}{T_\mathrm{eqm}} \times T_\mathrm{eqm} 
    = c_T \times T_1 \Delta \hat{u}_S (\Theta) \left ( \frac{\rh}{1\au} \right )^{-2} 
    \equiv c_T \Delta \hat{T}_1
    ~,
\end{equation*}
for a dimensionless temperature used in TPM, $ u = T / T_\mathrm{eqm} $ (and $ \Delta \hat{u}_S = \Delta \hat{T}_S / T_\mathrm{eqm} $). The $ u $ profile, and thus the $ \Delta \hat{u}_S $, depend only on the thermal parameter $ \Theta $ (and the spin vector; see Sect. \ref{meth-tpm}). The term on the right-hand side, $ \Delta \hat{T}_1 := T_1 \Delta u (\Theta) (\frac{\rh}{1\au})^{-2} $, is the $ \Delta T_S $ when the asteroid is at $ \rh $ but with $ c_T = 1 $ (e.g., $ A_B = 0 $ and $ \bar{\epsilon}_S = 1 $; see Eq. \ref{eq: c_T-def}). All values other than $ c_T $ are calculable once the coordinate $ (\rh,\, \Theta) $ is given, and so it is mathematically convenient to use  
\begin{equation}
  \frac{1}{c_T} \Delta \hat{T}_S > \frac{100\,\mathrm{K}}{c_T}
\end{equation}
rather than $ \Delta \hat{T}_S > 100\,\mathrm{K} $ to represent general asteroids.

Similarly, the second criterion is reformulated using
\begin{equation*}
  \frac{\Delta T}{\Delta \tau}
    = \frac{\Delta u T_\mathrm{eqm}}{P_\mathrm{rot} / \mathtt{nlon}}
    = c_T \Delta u(\Theta) T_1 \frac{\mathtt{nlon}}{60}  \left ( \frac{\rh}{1\au} \right )^{-2} 
       \left ( \frac{P_\mathrm{rot}}{1 \h} \right )^{-1} \, [\mathrm{K/min}]~,
\end{equation*}
where $ \Delta \tau = \frac{P_\mathrm{rot}}{\mathtt{nlon}}$ is the time bin and $ \mathtt{nlon} $ is the number of longitude bin ($ \mathtt{nlon} $ is fixed in the TPM simulation). $ \Delta u(\Theta) $ on the right-hand side is calculated numerically by taking the difference of the consecutive $ u $ array elements in the code ($ \mathtt{u}[1:] - \mathtt{u}[:-1] $). Therefore, $ \Delta u(\Theta) T_1 \left ( \frac{\rh}{1\au} \right )^{-2} $ is obtained for the coordinate grid $ (\rh,\, \Theta) $. $ \frac{\mathtt{nlon}}{60} $ is a constant for the conducted TPM. The second criterion is obtained by taking the maximum of the absolute values of the $ \Delta u $ array in the code, and multiplying it by the constants. Thus, it is more convenient to use 
\begin{equation}
  \frac{(P_\mathrm{rot} / 1\h)}{c_T} \left | \frac{d\hat{T}_S}{dt} \right | 
    > \frac{(P_\mathrm{rot} / 1\h)}{c_T} \times 2 \,\mathrm{K}/\min
\end{equation}
rather than $ \left | d\hat{T}_S/dt \right | > 2\,\mathrm{K}/\min $ to represent general asteroids.

To generate the plot in Fig. \ref{fig:theta-rh}, we set the grid of $ \Theta $ on a logarithmic scale: $ \Theta = 10^x $ for 50 uniformly-spaced $ x \in [-2, +2]$. The aspect angle is fixed to $ 90\degr $. First the equatorial temperature profile for these 50 models, $ u(\Theta) = T/T_\mathrm{eqm} $, are calculated (see Sect. \ref{meth-tpm}). Then the temperature is restored by multiplying $ T_\mathrm{eqm} $ at each $ \rh $ with $ c_T = 1 $. The contours are calculated for the re-formulated formulae, i.e., the contours are $ \frac{1}{c_T} \Delta \hat{T}_S $ and $ \frac{(P_\mathrm{rot} / 1\h)}{c_T} \left | \frac{d\hat{T}_S}{dt} \right | $ for a fictitious object of $ c_T = 1 $ and $ P_\mathrm{rot} = 1\h $.

For $ c_T \ll 1 $ ($ A_B \rightarrow 1 $), the first condition pushes the region for thermal fatigue to the lower left corner in Fig. \ref{fig:theta-rh}. This is expected because, if most sunlight is reflected, we need small thermal conduction ($ \Gamma \rightarrow 0 $) and higher insolation ($ \rh \rightarrow 0 $). Meanwhile, for a fast rotator ($ P_\mathrm{rot} \rightarrow 0 $), the second criterion requires a similar condition. This is also expected, because the temperature profile will be nearly constant at a given latitude for such an asteroid, and the temperature change rate will be very small unless it is close to the Sun and there is small thermal conduction.

\section{Code and data availability} \label{app: code}
All the scripts, source codes and developed packages used to generate datasets and plots in this work are available via the \textit{GitHub} service\footnote{https://github.com/ysBach/thermal\_radiation01}. As the model calculation may take a long time, the tables of $ Q_\mathrm{pr} $ and $ \bar{Q}_\mathrm{pr} $, as well as the refractive indices of olivine and magnetite, are available from the \textit{GitHub} repository. Also, the dataset generated by one of the notebooks in the \textit{GitHub} repository (\texttt{04-accplot.ipynb}), which is used for generating Fig. \ref{fig:size-size_all}, \ref{fig:rhdiammax}, and \ref{fig:phaeacceq}, is accessible via a separate \textit{figshare} service\footnote{https://doi.org/10.6084/m9.figshare.12044883}, due to its large size. The package \texttt{miepython} is also available via the \textit{GitHub} service\footnote{https://github.com/scottprahl/miepython; the specific version we used is  https://github.com/scottprahl/miepython/tree/0129cedb231c5e60170860b69cdb06eb26be5669}.

\end{appendix}
%%%%%%%%%%%%%%%%%%%%%%%%%%%%%%%%%%%%%%%%%%%%%%%%%%%%%%%%%
% END
%%%%%%%%%%%%%%%%%%%%%%%%%%%%%%%%%%%%%%%%%%%%%%%%%%%%%%%%%

\end{document}